\documentclass[prb,twocolumn]{revtex4}
\usepackage{amsfonts}
\usepackage{amsmath}
\usepackage{amssymb}
\usepackage{graphicx}
\begin{document}
\title{Magnetic oscillations of critical current in intrinsic Josephson-junction stacks}
\author{A. E. Koshelev}
\affiliation{Materials Science Division, Argonne National Laboratory, Argonne, Illinois 60439}
\date{\today }

\begin{abstract}
A key phenomenon related to the Josephson effect is oscillations of
different properties of superconducting tunneling junctions with
magnetic field. We consider magnetic oscillations of the critical
current in stacks of intrinsic Josephson junctions, which are
realized in mesas fabricated from layered high-temperature
superconductors. The oscillation behavior is very different from the
case of a single junction. Depending on the stack lateral size,
oscillations may have either the period of half flux quantum per
junction (wide-stack regime) or one flux quantum per junction
(narrow-stack regime). We study in detail the crossover between
these two regimes. Typical size separating the regimes is
proportional to magnetic field meaning that the crossover can be
driven by the magnetic field. In the narrow-stack regime the lattice
structure experiences periodic series of phase transitions between
aligned rectangular configuration and triangular configuration.
Triangular configurations in this regime are realized only in narrow
regions near magnetic-field values corresponding to integer number
of flux quanta per junction.
\end{abstract}
\maketitle

\section{Introduction}

Layered high-temperature superconducting materials, such as Bi$_{2}$Sr$_{2}
$CaCu$_{2}$O$_{x}$ (BSCCO), are composed of superconducting cuprate layers
coupled by Josephson interaction. This system possesses the Josephson effects
at the atomic scale (``Intrinsic Josephson Effect''). A rich spectrum of
classical \emph{dc} and \emph{ac} Josephson phenomena have been observed in
this system, see reviews \onlinecite{IJJReviews,YurgensReview}.

In a bulky superconductor the magnetic field applied along the layers
generates a triangular lattice of Josephson vortices. The anisotropy factor
$\gamma$ and the interlayer periodicity $s$ set the important field scale,
$B_{\mathrm{cr}}=\Phi_{0}/(2\pi\gamma s^{2})$ ($\sim$ 0.5 tesla for BSCCO).
When the magnetic field exceeds $B_{\mathrm{cr}}$ the Josephson vortices
homogeneously fill all layers.\cite{BulClemPRB91} Strong coupling between the
vortex arrays in neighboring layers mediated by the in-plane supercurrents
\cite{Inductive} determines the static and dynamic properties of the lattice.
Dynamic properties of the Josephson-vortex lattice in BSCCO have been
extensively studied by several experimental groups (see, e.g.,
Refs.\ \onlinecite{Lee,Hechtfischer,Latyshev,Ooi01}).

When an external transport current flowing across the layers exceeds
the \emph{critical current}, the Josephson vortex lattice starts to
move. In a homogeneous junction the critical current is determined
by interaction with the boundaries. The simplest and most known case
is a single small junction without inhomogeneities, where the field
dependence of the critical current is given by the Fraunhofer
dependence, $I_{c}(\Phi)\!=\!I_{c0}|\sin(\pi\Phi
/\Phi_{0})|/(\pi\Phi/\Phi_{0})$, with $\Phi$ being the magnetic flux
through the junction. Observation of this dependence has been
considered as an important confirmation of the \textit{dc} Josephson
effect.\cite{RowelPRL63} The same dependence is also expected for
the junction stack with the lateral size smaller than the Josephson
length.\cite{BulPRB92} In a single long junction the critical
current has the rather complicated field dependence due to multiple
coexistent states of the lattice.\cite{LongJosJunct}

In the previous paper \onlinecite{AEKPRB02} we considered the
behavior of the critical current for the dense Josephson-vortex
lattice in a homogeneous wide stack for which the critical current
is caused by interaction with the boundaries. We found that the
boundary induces an alternating deformation of the lattice.
Averaging out the rapid phase oscillations, we obtained that the
lattice deformation obeys the sine-Gordon equation and decays inside
superconductor at the typical length $L_{B}/\sqrt{8}$, which is
larger than the Josephson length, $\lambda_{J}\!=\!\gamma s$, and
increases proportional to the magnetic field, $L_{B}\!=\!\lambda_{J}
B/B_{\mathrm{cr}}$ .\cite{note-length} The stack is in the
wide-stack regime if its lateral width $L$ is larger than this
typical length. In this situation the surface deformation and the
total current flowing along the surface is uniquely determined by
the lattice position far away from the boundaries. The surface
current has oscillating dependence on the lattice displacement and,
due to the triangular-lattice ground state, the period of this
dependence is half the lattice spacing. The total current flowing
through the stack is given by the sum of two independent surface
currents flowing at the sample edges. The magnetic field determines
the magnitude of the maximum surface current (it is inversely
proportional to the field) and sets the phase shift between the
oscillating dependences of the two surface currents on the lattice
position. One can trace that, due to the half-lattice-spacing
periodicity of the surface current, a full period change of this
phase shift corresponds to the change of the magnetic flux through
one junction, $\Phi$, equal to the half flux quantum, $\Phi_{0}/2$.
As a consequence, the maximum current through the stack has
oscillating field dependence, which resembles the Fraunhofer
dependence: it has strong oscillations and overall $1/B$ dependence.
However, the period of these oscillations is two times smaller: it
corresponds to adding one flux quantum per two junctions and the
critical current has local maxima at $\Phi=k\Phi_{0}/2$.

Oscillations of the flux-flow voltage in BSCCO mesas at slow lattice
motion have been observed by Ooi \emph{et al.}\onlinecite{Ooi01} The
oscillations have the period of $\Phi_{0}/2$ per junctions and are
caused the size-matching effect described in the previous paragraph.
These oscillations have been reproduced by numerical
simulations.\cite{Machida03} More recently the flux-flow
oscillations in BSCCO mesas have been reproduced and studied in more
details by several experimental
groups.\cite{Kakeya05,ZhuPRB05,Urayama06,LatyshevPZhETF05} Similar
oscillations also have been observed in underdoped
YBa$_{2}$Ca$_{3}$O$_{6+x}$.\cite{NagaoPRB06} Size dependence of
oscillations has been systematically studied in Refs.\
\onlinecite{Kakeya05,Urayama06}. It was found that at smaller
lateral sizes and/or higher magnetic fields the crossover to the new
oscillation regime takes place, in which the period becomes
$\Phi_{0}$ per junction, as in a single junction.

Being motivated by recent experiments, in this paper we extend our
consideration to the regime when the junction size $L$ is comparable
with the length $L_{B}$ and the system crosses over from the
wide-stack to narrow-stack regime. As the length scale $L_{B}$
increases with the magnetic field, it also sets the field scale
$B_{L}= B_{\mathrm{cr} }L/\lambda_{J}$, at which this length becomes
of the order of the junction length $L$. Therefore for a junction of
a given size the crossover to the narrow-stack regime can be driven
by the magnetic field, as it was observed
experimentally.\cite{Kakeya05} $\Phi_{0}/2$-periodicity of the
critical-current oscillations holds until interactions between
surface deformations can be neglected. This interaction becomes
progressively stronger with decreasing the ratio $L/L_{B} $. Surface
deformations at $L>L_{B}$ can be described as partial sine-Gordon
solitons.\cite{SolitonNote} The relative sign of two solitons at the
opposite edges is determined by the magnetic field and the lattice
positions. At the integer-flux-quanta points $\Phi=k\Phi_{0}$ the
surface solitons always have the same sign and repel each other. As
a consequence, the amplitude of surface deformations drops and the
critical current decreases. At the half-integer-flux-quanta points
$\Phi=(k+1/2)\Phi_{0}$ situation is the opposite: the surface
solitons always have opposite signs and attract each other leading
to enhancement of the surface deformations and increase of the
critical current. Therefore the interaction between the surface
solitons leads to the crossover between the $\Phi_{0}/2$-periodic
oscillations of the critical current and $\Phi_{0}$-periodic
oscillations. This crossover occurs via suppression of the current
peaks at the points $\Phi=k\Phi_{0}$ and enhancement of the current
peaks at the points $\Phi=(k+1/2)\Phi_{0}$. Such behavior is
consistent with recent studies of the oscillations of the flux-flow
voltage in mesas with small lateral
sizes.\cite{Kakeya05,ZhuPRB05,Urayama06} The crossover in the
voltage oscillations also has been studied
numerically.\cite{ZhuPRB05,IrieOyaSST07}

In the region $L\sim L_{B}$ the lattice structure is determined by
competition between two energies: the interaction with boundaries
and the bulk shearing interaction between the Josephson-vortex
planar arrays in neighboring layers. The interaction with the
boundaries favors the aligned rectangular arrangement of the
Josephson vortices while the local shearing interaction favors the
triangular lattice. The boundary interactions decay slower with
increasing field then the shearing interaction and become dominating
at large fields. On the other hand, the boundary interaction energy
has oscillating field dependence and vanishes at the points
$\Phi=k\Phi_{0}$. At these points the shearing interaction is
relevant at any magnetic field. In addition, the interaction with
the boundaries is suppressed by the external current.

In the region $L\ll L_{B}$ the rectangular arrangement of vortices
is realized in most part of the phase space and the field dependence
of the critical current approaches the classical Fraunhofer
dependence. Two important deviations persist at all fields and
sizes: (i) Near the points $\Phi =k\Phi_{0}$ the phase transitions
to the triangular lattice always take place. Critical current at
these points never drops to zero and actually always has a small
local maximum; (ii) Away from the points $\Phi=k\Phi_{0}$ the
critical current is reached at the instability point of the
rectangular vortex lattice and it is always somewhat smaller than
the ``Fraunhofer'' value, $I_{c0}
|\sin(\pi\Phi/\Phi_{0})|/|\pi\Phi/\Phi_{0}|$. Therefore, we somewhat
revise the statement of Ref. \onlinecite{BulPRB92} that behavior of
a small stack is identical to a single small junction. The described
results has been summarized in a short paper published in the
conference proceedings.\cite{KoshCrete05} The static lattice
structures at different fields and sizes have been also studied
numerically in Refs.\ \onlinecite{MachidaPRL06} and
\onlinecite{IrieOyaSST07} and the results are in a qualitative
agreement with the described picture.

\begin{figure}[ptb]
\begin{center}
\includegraphics[width=3.4in]{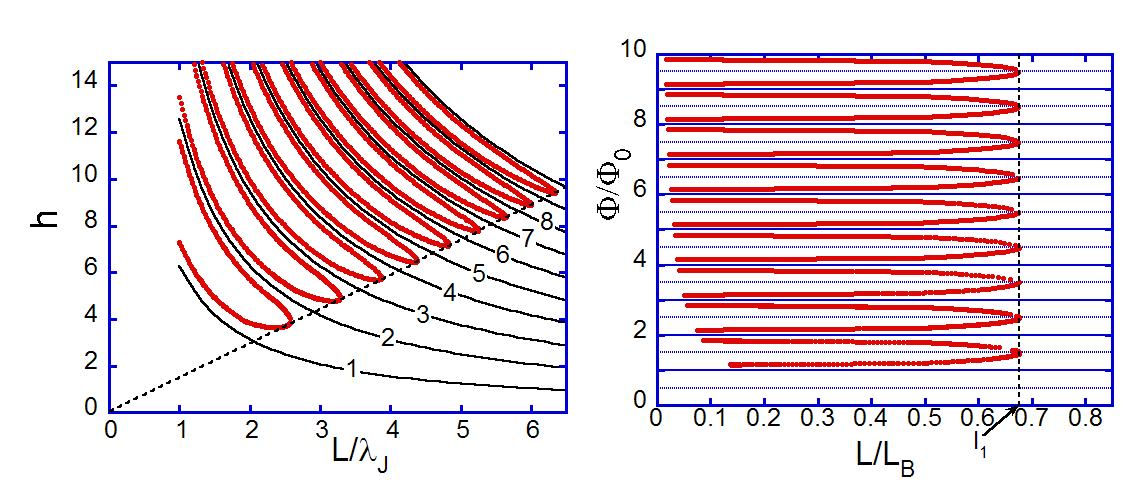}
\end{center}
\caption{\emph{Left plot:} Phase diagram of the Josephson-junction
stack in the coordinates [reduced junction size
$L/\lambda_{J}$]-[reduced field $h\!=\!B/B_{cr}$]. The solid lines
mark fields corresponding to the integer values of the magnetic flux
per junctions, $\Phi\!=\!k\Phi_{0}$. Dotted lines show boundaries of
the transitions into the rectangular lattice for ground state.
\emph{Right plot:} The same diagram in the coordinates [number of
flux quanta
per junction $\Phi/\Phi_{0}$]-[ratio $L/L_{B}\equiv L/(h\lambda_{J}) $].}
\label{Fig-h-LDiagram}
\end{figure}

To illustrate a general picture, we show in Fig.\ \ref{Fig-h-LDiagram} (left)
the ground-state phase diagram of the junction stack in the field-size plane,
where $h\!=\!B/B_{\mathrm{c}r}$ is the reduced magnetic field. Solid lines
mark the magnetic fields corresponding to integer values of the magnetic flux
per junction and dotted lines show boundaries of the rectangular-lattice
regions for zero current through the stack. As one can see, these regions
appear above the line $h=1.48L/\lambda_{J}$ and they are always bounded by the
integer-flux quanta lines. In the right plot the same diagram is replotted in
different coordinates, number of flux quanta per junction $\Phi/\Phi_{0}$ vs
ratio $L/L_{B}\equiv L/(h\lambda_{J})$, which controls the
wide-stack/narrow-stack crossover. It is interesting to note that in these
coordinates the diagram is periodic with respect to the magnetic flux through
the junction. 

The paper is organized as follows. In Section \ref{Sec:PhaseDist} we
outline derivation of the phase distribution in the case of
alternating from layer to layer solution. Averaging with respect to
the rapidly changing phase oscillations, we obtain equation and
boundary conditions for the slow lattice deformation. We also
express the lattice energy and current flowing through the stack via
this deformation. In Section \ref{Sec:Solut} we obtain and analyze
solution for the smooth phase in terms of elliptic integrals. We
found that the problem can be reduced to solution of three nonlinear
coupled equations for three unknowns, the boundary phases and
elliptic parameter. In Section \ref{Sec:Rect} we derive a criterion
for the transition into the rectangular-lattice state. In Appendix
\ref{App:LargeL} we consider weak finite-size effects in the
wide-stack regime and analytically compute exponentially small
finite-size corrections to the critical current, which break the
$\Phi_{0}/2$-periodicity of oscillations. In Section
\ref{Sec:Crossover} we present results of numerical analysis of the
crossover between the wide-stack and narrow-stack regimes with the
increasing magnetic field. We obtain the oscillation patterns of the
critical current for stacks with different lateral sizes and find
location of the rectangular-lattice regions in the current-field
plane. In Section \ref{Sec:NarrowStack} we reanalyze in detail the
narrow-stack regime using independent analytical approach. In
Section \ref{Sec:VoltOsc} we consider the voltage oscillations in
the case of slowly moving lattice and relate these oscillations with
the critical-current oscillations. We elaborate the recipe to
extract the anisotropy factor from the voltage oscillations.

\section{Phase distribution and energy of finite stack assuming alternating
solution \label{Sec:PhaseDist}}

We consider a Josephson-junction stack consisting of $N$ layers,
$N\gg 1$, with lateral size equal to $L$ in a magnetic field
$B>B_{\mathrm{c}r}$ applied along the layers. At high magnetic
fields one can neglect screening effects.
In this case the stack is described by the energy functional (per
layer and per unit length in the field direction) of the layer
phases $\varphi_n(x)$,
\begin{align}
E[\varphi_n]=&\frac{1}{N}\sum_{n}\int_{0}^{L}dx\left[
\frac{J}{2}\left( \frac{d\varphi_{n}}{dx}\right)^{2}\right.\nonumber\\
&\left.-E_{J}\cos\left(  \varphi_{n+1} -\varphi_{n}-\frac{2\pi
sB}{\Phi_{0}}x\right)  \right]  . \label{Energy}
\end{align}
where $J$ is the in-plane phase stiffness and $E_{J}$ is the
Josephson energy per unit area. To simplify derivations, we
introduce reduced coordinate, $u=x/\lambda_J$, with
$\lambda_J=\sqrt{J/E_J}$, reduced magnetic field, $h=2\pi s\lambda_J
B/\Phi_0$, and reduced energy $\mathcal{E}=E/(E_{J}\lambda_{J})$. We
also represent the phase variable in the form, which naturally
describes the dense triangular lattice in the bulk in the limits
$h\gg 1$ and $L/\lambda_J \gg h$, $\varphi_{n}(u)=\phi_{n}(u)+\alpha
n+\pi n(n-1)/2$, where the phases $\phi_{n}(u)$ are assumed to be
small and rapidly oscillating and the parameter $\alpha$ will
describe lattice displacement. The the reduced energy per one
junction and per unit length in the field direction can now be
represented as
\begin{align}
\mathcal{E}[\phi_{n}]  = &  \frac{1}{N}\sum_{n}\int_{0}^{L}du\left[
\frac{1}{2}\left(
\frac{d\phi_{n}}{du}\right)  ^{2}\right. \nonumber\\
-  &  \left.  \cos\left(  \phi_{n+1}-\phi_{n} -hu+\alpha+\pi
n\right) \right]. \label{RedEner}
\end{align}
The oscillating behavior is determined by the reduced parameter $hL$
which is directly related to the total magnetic flux through one
junction $\Phi=BLs$,
\[
hL=2\pi\Phi/\Phi_{0}.
\]
We consider the stack containing a very large number of junctions,
$N\gg 1$. This will allow us to focus on bulk behavior and neglect
c-axis boundary effects coming from the top and bottom junctions,
which give $1/N$ corrections to the bulk results. We will also not
consider potentially interesting ``parity effects'', small
differences between stacks containing odd and even number of
junctions, which have the same order.
Our results are also not influenced by possible perturbations of the
current distribution near the boundaries. Due to the large
anisotropy of the material, the ``bulk'' current distribution
usually is realized already in the second junction in the stack.

Following Ref.\ \onlinecite{AEKPRB02}, we will assume the
alternating phase distribution in the form
$\phi_{n}\!=\!(-1)^{n}\phi$. This distribution describes both the
deformed triangular lattice in the wide-stack regime and the
transition to the rectangular lattice in narrow-stack regime.
Substituting this presentation into Eq.\ (\ref{RedEner}), we
represent the energy functional as
\begin{equation}
\mathcal{E}(\alpha;\phi)\!=\!\int_{0}^{L}\!du\!\left[
\frac{1}{2}\left(  \! \frac{d\phi} {du}\!\right)
^{2}\!\!-\!\sin\left(  2\phi\right)  \sin\left(
\!-hu\!+\!\alpha\right)  \right]. \label{EnAlt}
\end{equation}
The lattice energy as function of displacement,
$\bar{\mathcal{E}}(\alpha)$, is determined by the minimum of the
functional $\mathcal{E}(\alpha;\phi(u))$ with respect to $\phi(u)$,
$\bar{\mathcal{E}}(\alpha)=\min_{\phi}\left[
\mathcal{E}(\alpha;\phi)\right] $. As the energy functional has a
symmetry property $\alpha\rightarrow\alpha+\pi$,
$\phi\rightarrow-\phi$, the energy $\bar{\mathcal{E}}(\alpha)$ is
$\pi$-periodic with respect to shift of $\alpha$,
$\bar{\mathcal{E}}(\alpha+\pi)=$ $\bar{\mathcal{E}}(\alpha)$.

The ground-state phase distribution $\phi$ obeys the following equation
\begin{equation}
\frac{d^{2}\phi}{du^{2}}+2\cos\left(  2\phi\right)  \sin\left(  -hu+\alpha
\right)  =0, \label{Phi-eq}
\end{equation}
which has to be solved with the boundary conditions
\begin{equation}
\frac{d\phi}{du}=0\text{, for }u=0,L. \label{Phi-Bound}
\end{equation}
In the limit $h\gg1$ further significant simplification is possible: we can
average out rapid phase oscillations and derive a simplified equation for the
smooth phase perturbation. We split the total phase into the smooth and
rapidly-oscillating components
\begin{equation}
\phi(u)=v(u)+\tilde{\phi}(u), \label{SplitPhase}
\end{equation}
where we assume $|\tilde{\phi}|$, $|dv/du|\ll1$. The maximum value of $v(u)$,
$v_{\mathrm{max}}=\pi/4$, corresponds to rectangular lattice. The
rapidly-oscillating phase by definition obeys the equation
\begin{equation}
\frac{d^{2}\tilde{\phi}}{du^{2}}+2\cos\left(  2v\right)  \sin\left(
-hu+\alpha\right)  =0, \label{EqOscPhase}
\end{equation}
which has the following approximate solution
\begin{equation}
\tilde{\phi}\approx\frac{2}{h^{2}}\cos\left(  2v\right)  \sin\left(
-hu+\alpha\right)  . \label{OscPhase}
\end{equation}
To the first order with respect to $\tilde{\phi}$, equation for $v(u)$ is
given by
\[
\frac{d^{2}v}{du^{2}}-4\tilde{\phi}\sin\left(  2v\right)  \sin\left(
-hu+\alpha\right)  =0.
\]
Substituting into this equation the oscillating phase (\ref{OscPhase}) and
averaging it with respect to rapid oscillations, we finally obtain the
sine-Gordon equation for the smooth phase
\begin{equation}
\frac{d^{2}v}{du^{2}}-\frac{2}{h^{2}}\sin\left(  4v\right)  =0.
\label{EqSmooth}
\end{equation}
Computing the derivatives of the rapid phase (\ref{OscPhase}) at the edges, we
also derive the boundary conditions for the smooth phase,
\begin{subequations}
\label{BoundSmooth}
\begin{align}
\frac{dv}{du}(0)  &  =\frac{2}{h}\cos\left(  2v_{0}\right)  \cos\left(
\alpha\right)  ,\label{BoundSmooth0}\\
\frac{dv}{du}(L)  &  =\frac{2}{h}\cos\left(  2v_{L}\right)  \cos\left(
-hL+\alpha\right)  \label{BoundSmoothL}
\end{align}
with $v_{0}\equiv v(0)$ and $v_{L}\equiv v(L)$. The local current density,
$j(u)=j_{J}\sin[\theta_{n,n+1}(u)]$ with $j_{J}$ being the maximum Josephson
current density, is determined by the gauge-invariant phase difference,
$\theta_{n,n+1}(u)\equiv\phi_{n+1}-\phi_{n}-hu+\alpha+\pi n$, which is related
to $v(u)$ as
\end{subequations}
\begin{align}
\theta_{n,n+1}  &  \approx-hu+\alpha+\pi n-(-1)^{n}2v\nonumber\\
&  -\frac{4}{h^{2}}\cos\left(  2v\right)  \sin\left(  -hu+\alpha+\pi n\right)
. \label{PhaseDiff}
\end{align}

Substituting the phase presentation (\ref{SplitPhase}) and
(\ref{OscPhase}) into the energy (\ref{EnAlt}) and averaging with
respect to the rapid oscillations, we derive the energy functional
in terms of the smooth phase $v$,
\begin{align}
\mathcal{E}(\alpha;v)  &  \approx\frac{1}{h}\left[  \sin\left(  2v_{0}\right)
\cos\left(  \alpha\right)  \!-\!\sin\left(  2v_{L}\right)  \cos\left(
-hL+\alpha\right)  \right] \nonumber\\
&  +\int_{0}^{L}du\left[  \frac{1}{2}\left(  \frac{dv}{du}\right)  ^{2}
-\frac{1+\cos4v}{2h^{2}}\right]  . \label{EnSmooth}
\end{align}
To shorten notations, we omitted the arguments $h$ and $L$ in
$\mathcal{E} (\alpha,h,L;v)$. Minimization of this energy functional
with respect to $v(u)$ gives the energy as a function of the lattice
shift $\alpha$, $\bar{\mathcal{E}}(\alpha)$. Minimum of this energy
with respect to $\alpha$ gives the ground state for given $h$ and
$L$. Higher-energy states at other values of $\alpha$ typically
carry a finite current. The total Josephson current flowing through
the stack is proportional to $d\bar{\mathcal{E}} /d\alpha$. Taking
derivative of the functional (\ref{EnSmooth}) with respect to
$\alpha$, assuming that at every $\alpha$ it is minimized with
respect to $v(u)$, we obtain
\begin{equation}
J(\alpha)\!=\!\frac{1}{h}\left[  -\!\sin\!\left(  2v_{0}\right) \sin
\!\alpha\!+\!\sin\!\left(  2v_{L}\right)  \sin\!\left(
\!-\!hL\!+\!\alpha \right)  \right]. \label{JosCurrSmooth}
\end{equation}
The unit of current in this equation is $j_{J}\lambda_{J}w$ where
$j_{J}$ is the Josephson-current density and $w$ is the junction
size in the field direction. An important consequence of this
equation is that nonzero current exists only if the surface
deformations $v_{0}$ and $v_{L}$ are finite. Further analysis is
based on Eqs.\ (\ref{EqSmooth}), (\ref{BoundSmooth}),
(\ref{EnSmooth}), and (\ref{JosCurrSmooth}).

\section{Solutions for smooth phase \label{Sec:Solut}}

A general solution of the sine-Gordon equation (\ref{EqSmooth}) can be found
in terms of elliptic integrals. From the first integral of
Eq.\ (\ref{EqSmooth}) we obtain
\begin{equation}
\frac{dv}{du}=\delta_{d}\sqrt{2}\frac{\sqrt{1/m-\cos^{2}\left(  2v\right)  }
}{h} \label{SmoothSolutDer}
\end{equation}
with $\delta_{d}\equiv\mathrm{sign}[dv/du]=\pm1$ and $m$ is the elliptic
parameter which has to be found from the boundary conditions. From this
equation we obtain implicit equation for deformation $v(u)$,
\begin{equation}
\int_{v_{0}}^{v}\frac{dv}{\sqrt{1/m-\cos^{2}\left(  2v\right)  }}=\delta
_{d}\sqrt{2}u/h. \label{SmoothSolut}
\end{equation}
To rewrite this equation using standard elliptic integrals, we
introduce a new variable $\varphi$,
\begin{equation}
\varphi=\pi/2+2v. \label{ElliptVar}
\end{equation}
This variable has its own physical meaning: it describes \emph{the
alternating deformation of the interlayer phase difference with
respect to the rectangular-lattice state}. In particular,
$\varphi=0$ corresponds to the rectangular lattice. Using these
variables, we can rewrite Eq.\ (\ref{SmoothSolut}) as
\begin{equation}
\sqrt{m}\left[  F(\varphi,m)-F(\varphi_{0},m)\right]  =
\delta_{d}\sqrt{8}L/h \label{Solut_phi_u}
\end{equation}
where
\[
F(\varphi,m)\equiv\int_{0}^{\varphi}\frac{dx}{\sqrt{1-m\sin^{2}x}}
\]
is the incomplete elliptic integral of the first
kind.\cite{Abramovitz}

In the limit of very large $L$ the deformation $v$ has to vanish far
away from edges meaning that $m\rightarrow1$. For finite-size
junctions, depending on $hL$ and $\alpha$, the solution $v(u)$ can
be either monotonic or nonmonotonic. For the nonmonotonic solution
the derivative $dv/du$ and the parameter $\delta_{d}$ change sign
inside. The monotonic solution can either change sign
($v_{0}v_{L}<0$, $m<1$) or not ($v_{0}v_{L}>0$). For large $L$ the
monotonic solution corresponds to the two surface partial solitons
of the same sign and the nonmonotonic case corresponds to the
surface solitons of opposite signs. A mathematical structure of
solutions for these two cases is different and we will consider them
separately.

First, we find some general relations between the boundary phases
and the parameter $m$. From the boundary conditions\
(\ref{BoundSmooth0}), (\ref{BoundSmoothL}), and \ Eq.\
(\ref{SmoothSolutDer}) we obtain equations
\begin{subequations}
\label{BoundCondSol}
\begin{align}
\delta_{0}\sqrt{1/m-\cos^{2}\left(  2v_{0}\right)  }  &  =\sqrt{2}\cos\left(
2v_{0}\right)  \cos\left(  \alpha\right)  ,\label{BoundCondSol0}\\
\delta_{L}\sqrt{1/m\!-\!\cos^{2}\!\left(  2v_{0}\right)  }  &
\!=\!\sqrt{2}\cos\!\left(
2v_{L}\right)  \!\cos\!\left(  hL\!-\!\alpha\right)  \label{BoundCondSolL}
\end{align}
\end{subequations}
with $\delta_{0}\!=\!\delta_{d}(0)$ and $\delta_{L}=\delta_{d}(L)$
(for monotonic solution $\delta_{0}\!=\!\delta_{L}$ and for
nonmonotonic solution $\delta _{0}=-\delta_{L}$). As
$|v_{0,L}|\!<\!\pi/4$, the inequality $\cos(2v_{0,L})\!>\!0$ always
holds meaning that $\delta_{0}\!=\!\mathrm{sign}[\cos\alpha]$ and
$\delta_{L}\!=\!\mathrm{sign}[\cos\left(  hL-\alpha\right)  ]$. From
the conditions (\ref{BoundCondSol}) we obtain the following results
for the boundary deformations
\begin{subequations}
\begin{align}
\cos\left(  2v_{0}\right)   &  =\sqrt{\frac{1/m}{1+2\cos^{2}\left(
\alpha\right)  }}\label{cosv0}\\
\cos\left(  2v_{L}\right)   &  =\sqrt{\frac{1/m}{1+2\cos^{2}\left(
hL-\alpha\right)  }} \label{cosvL}
\end{align}
\end{subequations}
or, in terms of variable $\varphi$ (\ref{ElliptVar}),
\begin{subequations}
\label{sin_phi_mon}
\begin{align}
\sin\varphi_{0}  &  =\sqrt{\frac{1/m}{1+2\cos^{2}\left(  \alpha\right)  }
},\label{sin_phi0_mon}\\
\sin\varphi_{L}  &  =\sqrt{\frac{1/m}{1+2\cos^{2}\left(
hL-\alpha\right)  }}.
\label{sin_phiL_mon}
\end{align}
\end{subequations}

\begin{figure}[ptb]
\begin{center}
\includegraphics[width=3.4in]{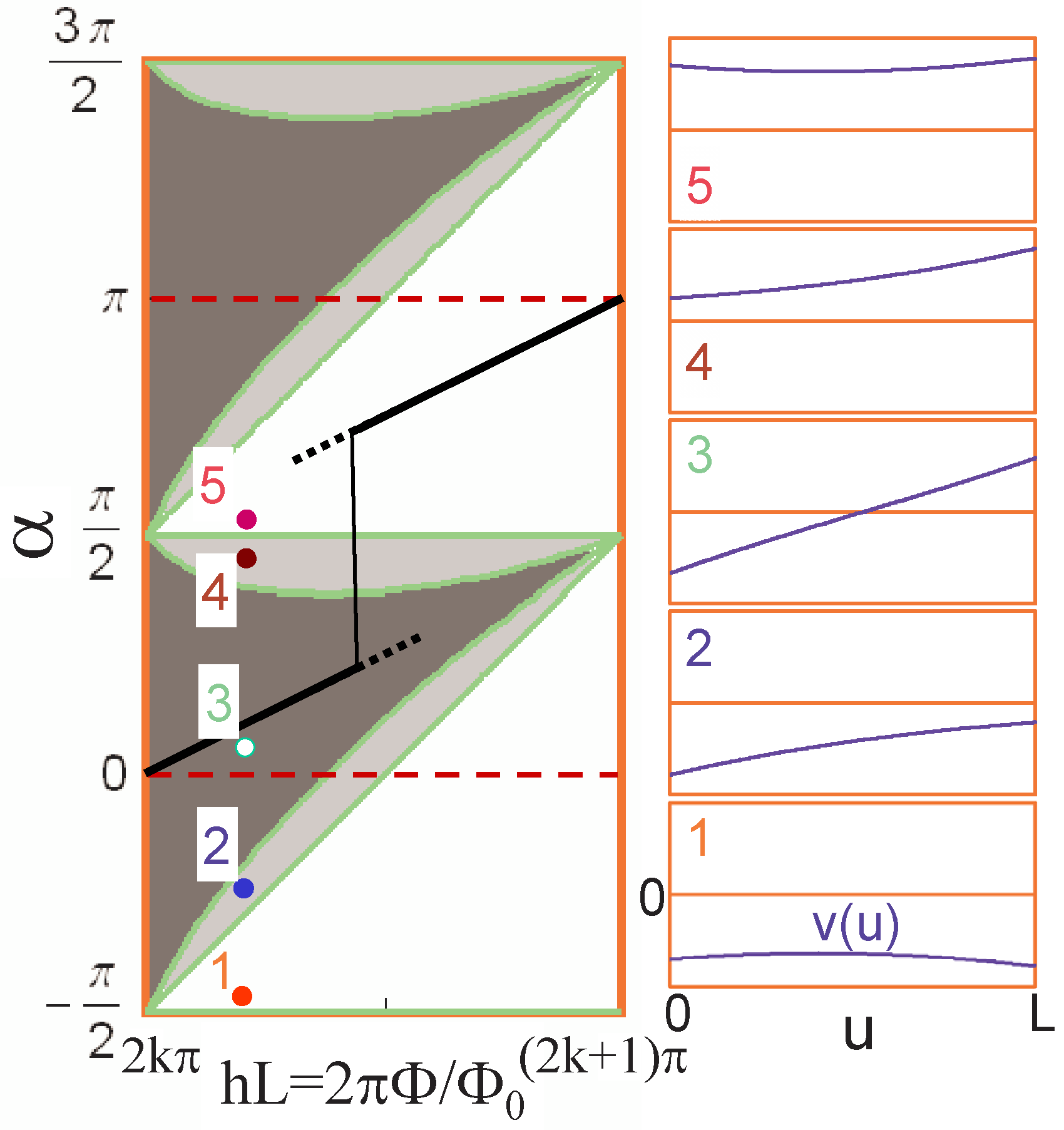}
\end{center}
\caption{Regions of different types of solution for the smooth alternating
deformation $v(u)$ in the plane $hL$-$\alpha$ for $L/h\sim1$. Typical
solutions are illustrated for five marked points. Dark grey color marks
regions of monotonic solution changing sign (3), light grey color marks
regions of monotonic solution not changing sign (2 and 4), and white regions
correspond to nonmonotonic solution (1 and 5). Light-grey regions shrink with
increasing $L$. The black line illustrates the location of the ground state.}
\label{Fig-SolutionDiagram}
\end{figure}

From these considerations one can conclude that the monotonic
solution realizes for all $\alpha$'s for $hL=2k\pi$ (magnetic flux
per junction equals to integer number of flux quanta,
$\Phi=k\Phi_{0}$) and the nonmonotonic solution realizes for
$hL=\left(  2k+1\right)  \pi$ (magnetic flux equals to half-integer
number of flux quanta, $\Phi=\left(  k+1/2\right)  \Phi_{0}$). If
the magnetic flux through one junction is not close to these special
values then the solution changes from monotonic to nonmonotonic
depending on the lattice phase shift $\alpha$. Location of different
types of solution depending on $hL$ and $\alpha$ is illustrated in
Fig.\ \ref{Fig-SolutionDiagram}.

One can distinguish two important special cases corresponding to
symmetric locations of the lattice with respect to the boundaries.
The first case, $\alpha=hL/2+\pi k$, $v_{L}=-v_{0}$, corresponds to
the monotonic solution and the second case,
$\alpha\!=\!hL/2\!+\!\pi/2\!+\!\pi k$, $v_{L}\!=\!v_{0}$,
corresponds to the nonmonotonic solution. The energy always has
extremums at these values of $\alpha$. Moreover, a detailed study
shows that the ground state is always realized at one of these
values of $\alpha$. At large $L$ the system switches between these
locations in the vicinity of the points $hL=(2k+1/2)\pi$, as it is
illustrated in Fig.\ \ref{Fig-SolutionDiagram}. In the vicinity of
these switching points the energy has minima at the both values of
$\alpha$.

In general, the conditions (\ref{cosv0}) and (\ref{cosvL}) are not
sufficient to determine signs of the edge deformations $v_{0}$ and
$v_{L}$. In the limit of large $L$ the deformation $v(u)$ has to
decay from the edges leading to relations
$\mathrm{sign}[v_{0}]=-\delta_{0}=-\mathrm{sign}[\cos\alpha]$ and
$\mathrm{sign}[v_{L}]=\delta_{L}=\mathrm{sign}[\cos\left(
hL-\alpha\right) ]$. In this case we also obtain conditions
\begin{align*}
\tan (2v_{0}) &  =-\delta_{0}\sqrt{m\left(  2\cos^{2}\left(
\alpha\right)
+1\right)  -1},\\
\tan (2v_{L}) &  =\delta_{L}\sqrt{m\left(  2\cos^{2}\!\left(
hL\!-\!\alpha\right)
+1\right)  -1}
\end{align*}
which fix signs of $v_{0}$ and $v_{L}$. For large values of $L/h$ monotonic
solution typically changes sign inside. However for finite $L$ there are
intermediate regions exist located near lines $\alpha=\pi/2+\pi k$ and
$\alpha=hL+\pi/2+\pi k$ where solution is still monotonic but does not change
sign. We now proceed with analyzing separately the monotonic and nonmonotonic
solutions. 

\subsection{Monotonic solution}

The monotonic solutions realize for ranges of $\alpha$ where
$\cos\alpha \cos\left(  hL-\alpha\right)  >0$ (grey regions in Fig.\
\ref{Fig-SolutionDiagram}). For such solution we obtain from Eq.\
(\ref{SmoothSolut}) relation connecting the parameter $m$ with the
boundary deformations
\begin{equation}
\int_{v_{0}}^{v_{L}}\frac{dv}{\sqrt{1/m-\cos^{2}\left(  2v\right)  }
}=\mathrm{sign}[\cos\alpha]\sqrt{2}L/h. \label{MonotRelat}
\end{equation}
Using previously introduced variable $\varphi$ (\ref{ElliptVar}), we
can rewrite Eq.\ (\ref{MonotRelat}) via elliptic integrals as
\begin{equation}
\sqrt{m}\left[  F(\varphi_{L},m)-F(\varphi_{0},m)\right]  =\mathrm{sign}
[\cos\alpha]\sqrt{8}L/h \label{Eq_m_monot}
\end{equation}
This equation together with boundary conditions
(\ref{sin_phi0_mon}), and (\ref{sin_phiL_mon}) have to be solved to
find the three unknown constants $\varphi_{0}$, $\varphi_{L}$, and
$m$, which completely determine the solution. The boundary
deformations $v_{0}\!=\!(\varphi_{0}\!-\!\pi/2)/2$ and
$v_{L}\!=\!(\varphi_{L}\!-\!\pi/2)/2$ may have either the same sign
or opposite signs. In Appendix \ref{App:same-sign} we find the
boundary separating these two types of solution (boundaries between
dark-grey regions and light-grey regions in Fig.\
\ref{Fig-SolutionDiagram}).

The energy (\ref{EnSmooth}) and Josephson current
(\ref{JosCurrSmooth}) can be represented as
\begin{align}
\mathcal{E}  &  =-\frac{1}{h}\left[  \cos\left(  \varphi_{0}\right)
\cos\left(  \alpha\right)  -\cos\left(  \varphi_{L}\right)  \cos\left(
hL-\alpha\right)  \right] \nonumber\\
&  +\frac{1}{\sqrt{2m}h}|E(\varphi_{L},m)-E(\varphi_{0},m)|-\frac{L}{mh^{2}
},\label{EnMonot}\\
J(\alpha)  &  =\frac{1}{h}\left[  \cos\left(  \varphi_{0}\right)  \sin
\alpha\!-\!\cos\left(  \varphi_{L}\right)  \sin\left(  -hL+\alpha\right)
\right]  , \label{CurrMonot}
\end{align}
where $E(\varphi,m)\equiv\int_{0}^{\varphi}\sqrt{1-m\sin^{2}\!x}\
dx$ is the incomplete elliptic integral of the second
kind.\cite{Abramovitz}

\subsection{Nonmonotonic solution}

The solution is nonmonotonic in the regions given by $\cos\alpha\cos\left(
hL-\alpha\right)  <0$ (white regions in Fig.\ \ref{Fig-SolutionDiagram}). In
this case the function $v(u)$ has extremum at some point $u=u_{m}$ so that we
can rewrite Eq.\ (\ref{SmoothSolutDer}) as
\begin{align*}
\frac{dv}{du} &  =\delta_{d}\sqrt{2}\frac{\sqrt{1/m-\cos^{2}\left(  2v\right)
}}{h},\ \text{for }u<u_{m},\\
\frac{dv}{du} &  =-\delta_{d}\sqrt{2}\frac{\sqrt{1/m-\cos^{2}\left(
2v\right)  }}{h},\ \text{for }u_{m}<u<L
\end{align*}
with $\delta_{d}\!=\!\mathrm{sign}[\cos\alpha]$. As $dv/du\!=\!0$ at
$u\!=\!u_{m}$, we have $m\!=\!1/\cos^{2}\!\left(  2v_{m}\right)
\!>\!1$ with $v_{m}\equiv v(u_{m})$. Integrating the first equation
from $0$ to $u_{m}$ and the second equation from $L$ to $u_{m}$, we
obtain
\begin{align*}
\int_{v_{0}}^{v_{m}}\frac{dv}{\sqrt{1/m-\cos^{2}\left(  2v\right)  }} &
=\delta_{d}\sqrt{2}u_{m}/h,\\
\int_{v_{L}}^{v_{m}}\frac{dv}{\sqrt{1/m-\cos^{2}\left(  2v\right)  }} &
=\delta_{d}\sqrt{2}\left(  L-u_{m}\right)  /h.
\end{align*}
Adding these two equations, we obtain equation connecting $v_{m}$ with the
boundary deformations $v_{0}$ and $v_{L}$
\begin{align}
&  \int_{v_{0}}^{v_{m}}\frac{dv}{\sqrt{\cos^{2}\left(  2v_{m}\right)
-\cos^{2}\left(  2v\right)  }}\nonumber\\
&  +\int_{v_{L}}^{v_{m}}\frac{dv}{\sqrt{\cos^{2}\left(  2v_{m}\right)
-\cos^{2}\left(  2v\right)  }}=\delta_{d}\sqrt{2}L/h.\label{NonMon-vm-Eq}
\end{align}
This equation together with boundary conditions (\ref{cosv0}) and
(\ref{cosvL}), represents the full system for determination of three
unknown constants $v_{0}$, $v_{L}$, and $m\!=\!1/\cos^2\!(2v_{m}) $.
To rewrite these equations in terms of elliptic functions, we again
transfer to variable (\ref{ElliptVar}). Then equation
(\ref{NonMon-vm-Eq}) can be rewritten as
\begin{equation}
\sqrt{m}\left[  2F(\varphi_{m},m)\!-\!F(\varphi_{0},m)\!-\!F(\varphi
_{L},m)\right]  \!=\!\delta_{d}\frac{\sqrt{8}L}{h}.\label{Eq_m_nonmon}
\end{equation}
with $\varphi_{m}=\pi/2+2v_{m}$. The boundary conditions in terms of
$\varphi_{0}$ and $\varphi_{L}$ are again given by Eqs.\ (\ref{sin_phi0_mon})
and (\ref{sin_phiL_mon}). The elliptic parameter is related to $\varphi_{m}$
as $m=1/\sin^{2}\varphi_{m}$ leading to the following relation, $F(\varphi
_{m},m)=K(1/m)/\sqrt{m}$. The energy for nonmonotonic solution can be
represented as
\begin{align}
\mathcal{E} &  =-\frac{1}{h}\left[  \cos\varphi_{0}\cos\left(  \alpha\right)
-\cos\varphi_{L}\cos\left(  -hL+\alpha\right)  \right]  \nonumber\\
\!+ &  \!\frac{1}{\sqrt{2m}h}\!\left\vert 2E(\varphi_{m},\!m)\!-\!E(\varphi
_{0},\!m)\!-\!E(\varphi_{L},\!m)\right\vert \!-\!\frac{L}{mh^{2}
}\label{EnNonMon}
\end{align}
and the Josephson current is again given by Eq.\ (\ref{CurrMonot}). One can
easily check that the nonmonotonic solution matches the monotonic solution at
the boundaries. For example, for $\cos\alpha=0$ the extremum is located at the
boundary $u=0$. In this case we have $\phi_{0}=\phi_{m}=\arcsin(1/\sqrt{m})$
(or $\pi-\arcsin(1/\sqrt{m})$) and Eq.\ (\ref{Eq_m_nonmon}) coincides with
Eq.\ (\ref{Eq_m_monot}).

\subsection{Alternative presentation of equations via elliptic integrals}

Using known relations for the elliptic integrals
\begin{subequations}
\begin{align}
F(\varphi,m)  &  \!=\!\frac{1}{\sqrt{m}}F(\psi,1/m)\ \text{with
}\sin
\psi\!=\!\sqrt{m}\sin\varphi,\label{ElliptRelF}\\
E(\varphi,m)  &  \!=\!\frac{1}{\sqrt{m}}\left[
(1\!-\!m)F\left(\psi,1/m
\right)\!+\!mE\left(\psi,1/m\right)\right]  , \label{ElliptRelE}
\end{align}
\end{subequations}
valid for $\varphi<\pi/2$ and $\sin\varphi<1/\sqrt{m}$, one can
rewrite equations (\ref{Eq_m_monot}) and (\ref{sin_phi_mon}) for the
case of the same-sign monotonic solution in the following
alternative form
\begin{subequations}
\begin{align}
F\left(  \psi_{L},\tilde{m}\right) & -F\left(
\psi_{0},\tilde{m}\right)
=\mathrm{sign}[\cos\alpha]\sqrt{8}L/h,\label{mon_psi_m}\\
\sin\psi_{0}  &  =\frac{1}{\sqrt{1+2\cos^{2}\left(  \alpha\right)  }
},\label{mon_psi0}\\
\sin\psi_{L}  &  =\frac{1}{\sqrt{1+2\cos^{2}\left(  hL-\alpha\right)
}},
\label{mon_psiL}
\end{align}
\end{subequations}
with $\tilde{m}\equiv1/m$. Correspondingly, the energy
(\ref{EnMonot}) in this representation is given by
\begin{align}
\mathcal{E}  &  \!=\!-\!\frac{1}{h}\left\vert
\sqrt{1\!-\!\tilde{m}\sin ^{2}\!\psi_{0}\!} \cos \alpha
\!-\!\sqrt{1\!-\!\tilde{m} \sin^{2}\!\psi_{L}\!}\cos\!\left(
hL\!-\!\alpha\right) \!\right\vert
\nonumber\\
&  +\frac{1}{\sqrt{2}h}\left\vert E\left(  \psi_{0},\tilde{m}\right)
-E(\psi_{L},\tilde{m})\right\vert -\frac{\left(  2-\tilde{m}\right)
L}{h^{2} }.
\end{align}

In the case of nonmonotonic solution, using relations
(\ref{ElliptRelF}) and (\ref{ElliptRelE}) for the elliptic
integrals, one can rewrite equation (\ref{Eq_m_nonmon}) in the
following equivalent form
\begin{equation}
2K\left(  \tilde{m}\right)  -F\left(  \psi_{0},\tilde{m}\right)
-F\left(
\psi_{L},\tilde{m}\right)  =\delta_{d}\sqrt{8}L/h. \label{nonmon_psi_m}
\end{equation}
where $\tilde{m}\equiv1/m$, $\psi_{0}$ and $\psi_{L}$ are given by
Eqs.\ (\ref{mon_psi0}) and (\ref{mon_psiL}), and represent the
energy as
\begin{align}
\mathcal{E}  &  =-\frac{1}{h}\left[
\sqrt{1-\tilde{m}\sin^{2}\psi_{0}
}\left\vert \cos \alpha  \right\vert \right. \nonumber\\
+  &  \left.  \sqrt{1-\tilde{m}\sin^{2}\psi_{L}}\left\vert
\cos\left(
hL-\alpha\right)  \right\vert \right] \nonumber\\
+  &  \frac{1}{\sqrt{2}h}\left\vert 2E\left(  \tilde{m}\right)
\!-\!E\left( \psi_{0},\tilde{m}\right)  \!-\!E\left(
\psi_{L},\tilde{m}\right)
\right\vert \!-\!\frac{(2\!-\!\tilde{m})L}{h^{2}}. \label{nonmon_En_psi}
\end{align}

This representation is especially useful in the case of large $m$
(small $\tilde{m}$). In particular, it will allow us to study
transition to the rectangular-lattice state corresponding to the
limit $m\rightarrow\infty$, which we will consider in the next
section.

\section{Transition to the rectangular lattice \label{Sec:Rect}}

An important particular case is the solution of Eq.\ (\ref{EqSmooth})
corresponding to rectangular lattice, $v=\pm\pi/4$ or $\varphi=0,\pi$. This
case corresponds to the limit  $m\rightarrow\infty$ ($\tilde{m}\rightarrow0$).
The energy of the rectangular lattice coincides with the well-known result for
a single junction
\begin{equation}
\mathcal{E}_{\mathrm{rect}}(\alpha)=-\frac{2}{h}\sin\left(  \frac{hL}
{2}\right)  \sin\left(  \alpha-\frac{hL}{2}\right)  \label{En_rect}
\end{equation}
and has minimum $\mathcal{E}_{\mathrm{rect}}=-2\left\vert \sin\left(
hL/2\right)  \right\vert /h$ at $\alpha=hL/2+\delta\pi/2$ with $\delta
=\mathrm{sign}\left[  \sin\left(  hL/2\right)  \right]  $.

To find condition for the transition to the rectangular lattice, we take the
limit $\tilde{m}\rightarrow0$ in Eq.\ (\ref{nonmon_psi_m}) for nonmonotonic
solution. Using relations $K(0)=\pi/2$ and $F(\psi,0)=\psi$, we obtain
\begin{equation}
\pi-\psi_{0}-\psi_{L}=\sqrt{8}L/h, \label{Condit-Rect}
\end{equation}
where $\psi_{0}$ and $\psi_{L}$ are given by Eqs.\ (\ref{mon_psi0}) and
(\ref{mon_psiL}). Using these definitions, the condition for the rectangular
lattice can be rewritten in an explicit form as
\begin{equation}
\frac{\sqrt{2}\left(  |\cos\left(  hL-\alpha\right)  |+|\cos\alpha|\right)
}{\sqrt{\left(  1\!+\!2\cos^{2}\alpha\right)  \left(  1\!+\!2\cos^{2}\left(
hL\!-\!\alpha\right)  \right)  }}\!<\!\sin\left(  \! \frac{\sqrt{8}L}
{h}\!\right)  . \label{Condit-Rect-Expl}
\end{equation}
This equation gives the transition criterion in general case, including the
current-carrying states. In particular, the rectangular lattice gives a local
energy minimum at $\alpha=hL/2+\pi/2$ in the regions $|hL/2\pi-(k+1/2)|<1/4$
if the inequality
\begin{equation}
\left\vert \sin\left(  hL/2\right)  \right\vert <\tan\left(  \sqrt
{2}L/h\right)  /\sqrt{2} \label{Cond-Rect-Ground}
\end{equation}
is satisfied. The rectangular lattice first appears in the ground
state at points $hL=(k+1/2)2\pi$ for $L/h\leq l_{1}=\arctan\left(
\sqrt{2}\right) /\sqrt{2}\approx0.675$. This value is marked in the
right plot of Fig.\ \ref{Fig-h-LDiagram}.

\section{Wide-stack/narrow-stack crossover \label{Sec:Crossover}}

\begin{figure}[ptb]
\begin{center}
\includegraphics[width=2.4in]{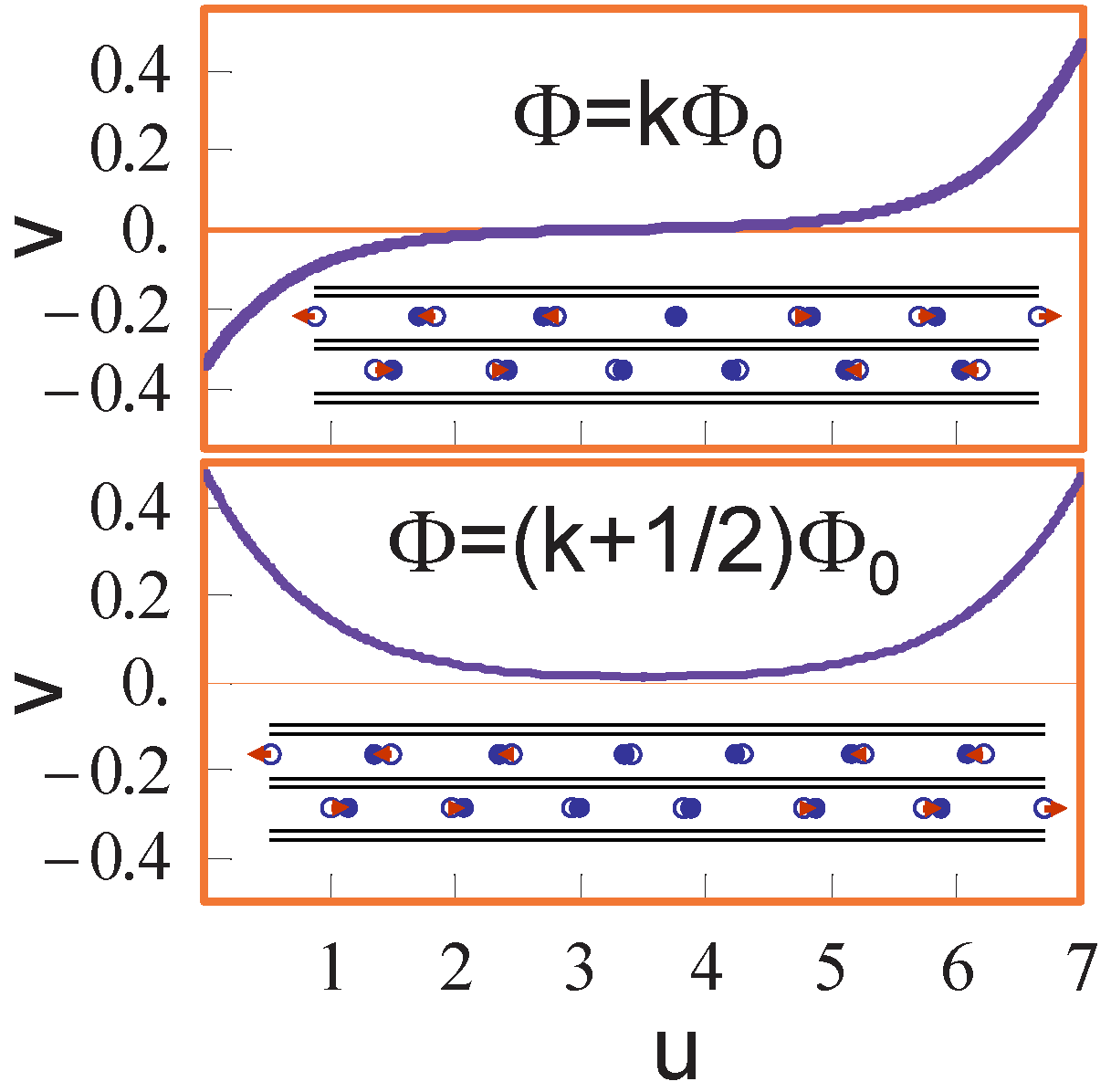}
\end{center}
\caption{Typical lattice deformations at large $L$ for
$\Phi=k\Phi_{0}$ (upper plot) and $\Phi=(k+1/2)\Phi_{0}$ (lower
plot). In the former case the surface partial solitons have the same
sign and repel each other while in the latter case they have the
opposite signs and attract each other. The insets in both plots
illustrate corresponding displacement fields of the Josephson-vortex
lattice in the two neighboring layers.}
\label{Fig-Solutons}
\end{figure}

In this section we investigate in detail the crossover between the
wide-stack and narrow-stack regimes. As this crossover is driven by
the reduced parameter $h/L$, for a junction with size $L$ the
crossover takes place with increasing magnetic field at
size-dependent field $B_{L}\!=\!L\Phi _{0}/(2\pi\gamma^{2}s^{3})$.

At large $L$, $L\gg h$ or $B\ll B_L$, the smooth alternating
deformation has solutions in the form of two isolated surface
solitons.\cite{SolitonNote} The monotonic solution corresponds to
the solitons of the same sign and the nonmonotonic solution
corresponds to the solitons of opposite signs, as it is illustrated
in Fig.\ \ref{Fig-Solutons}. If one neglects the interaction between
the solitons then the relative sign of surface soliton has no
importance and the total Josephson current is given by the sum of
two independent surface currents, which do not depend on the soliton
signs.\cite{AEKPRB02} As a consequence, the product $hJ_{c}$ has
periodicity of half flux-quantum per junction and reaches maxima at
$hL\!=\!\pi j$ ($\Phi\!=\!j\Phi_{0}/2$) with $hJ_{c}\approx 1.035$.
At finite $L$ the interaction between the surface solitons disturbs
such periodicity. At large $L$ one can derive analytically
corrections to the infinite-$L$ results, see Appendix
\ref{App:LargeL} for details. In particular, near the maxima
$hL\!=\!\pi j$ ($\Phi\!=\!j\Phi_{0}/2$), we find the finite-size
correction,
\begin{equation}
\delta J_{c}(h,\pi j)\approx-1.544\frac{(-1)^{j}}{h}\exp\left(
-\frac
{\sqrt{8}L}{h}\right)  , \label{CritCurrMaxLargeL}
\end{equation}
As we can see, the finite-size effects increase the critical current
maxima at $\Phi\!=\!(k\!+\!1/2)\Phi_{0}$ ($j\!=\!2k+1$) and reduce
the critical current maxima at $\Phi\!=\!k\Phi_{0}$ ($j\!=\!2k$). In
the wide-stack regime, however, these corrections are exponentially
small, which explains nice $\Phi_{0}/2$-periodic oscillation of the
flux-flow voltage observed in this regime.\cite{Ooi01,Kakeya05}

\begin{figure}[ptb]
\begin{center}
\includegraphics[width=3.4in]{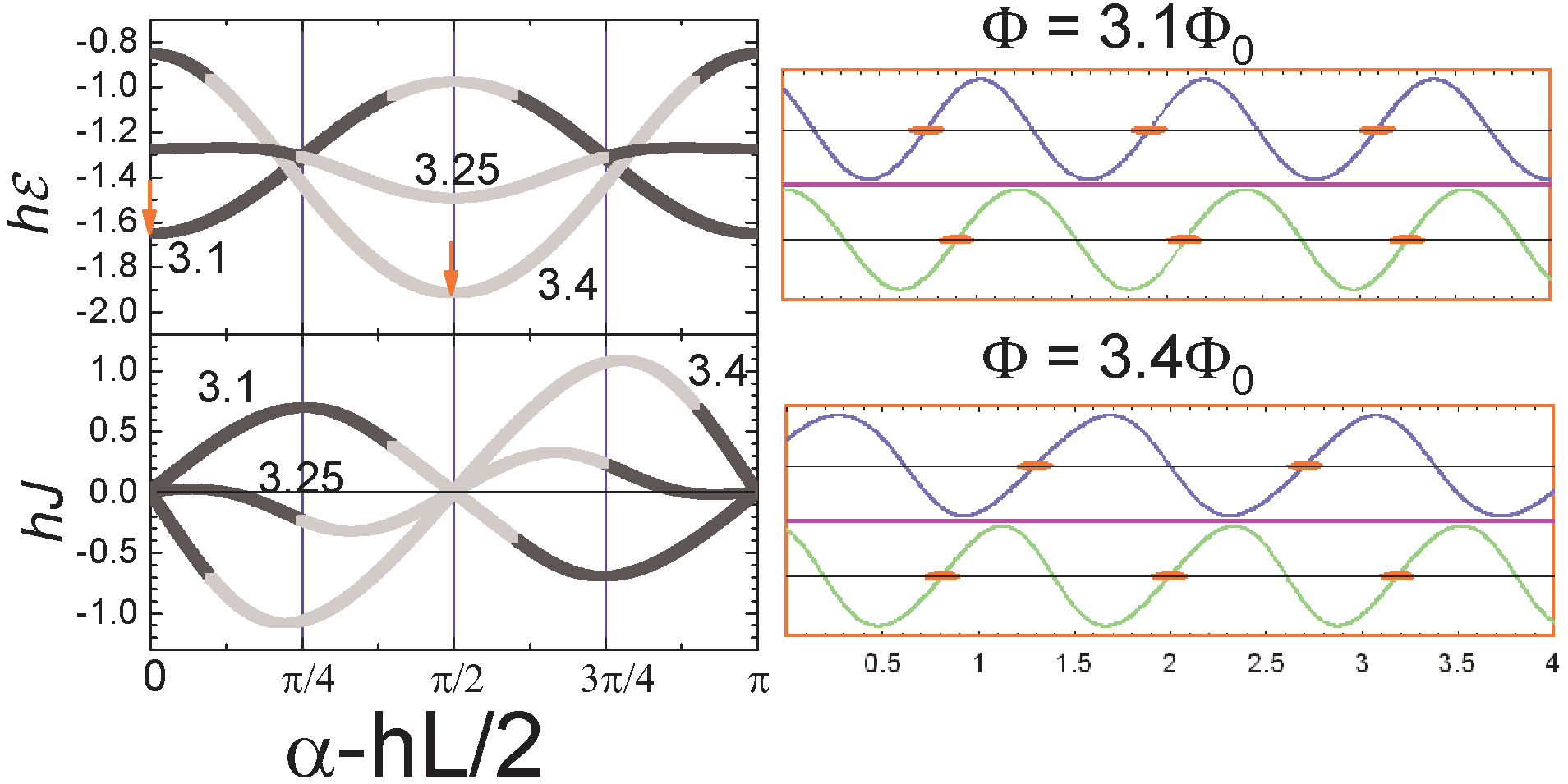}
\end{center}
\caption{The representative dependences of the energy (upper left
plot) and Josephson current (lower left plot) on parameter $\alpha$
for $L=4$ and three values of flux per junction $\Phi/\Phi_{0}=3.1$,
$3.25$, and $3.4$ within one oscillation period (the curves are
marked by these values) corresponding to the magnetic fields
$h=4.87$, $5.105$, and $5.34$. The parameters correspond to the
crossover region $L/h\!\sim\!1$. Dark grey marks region of monotonic
solution and light grey marks region of nonmonotonic solution.
Arrows in the energy plot mark values of $\alpha$ corresponding to
ground state. Pictures in the right column illustrate structure of
ground states at $\Phi/\Phi_{0}=3.1$ and $3.4$. The lines show
oscillating z-axis currents in neighboring layers
and ellipses mark centers of the Josephson vortices.}
\label{Fig-L4alpha}
\end{figure}\begin{figure}[ptb]
\begin{center}
\includegraphics[width=3.4in]{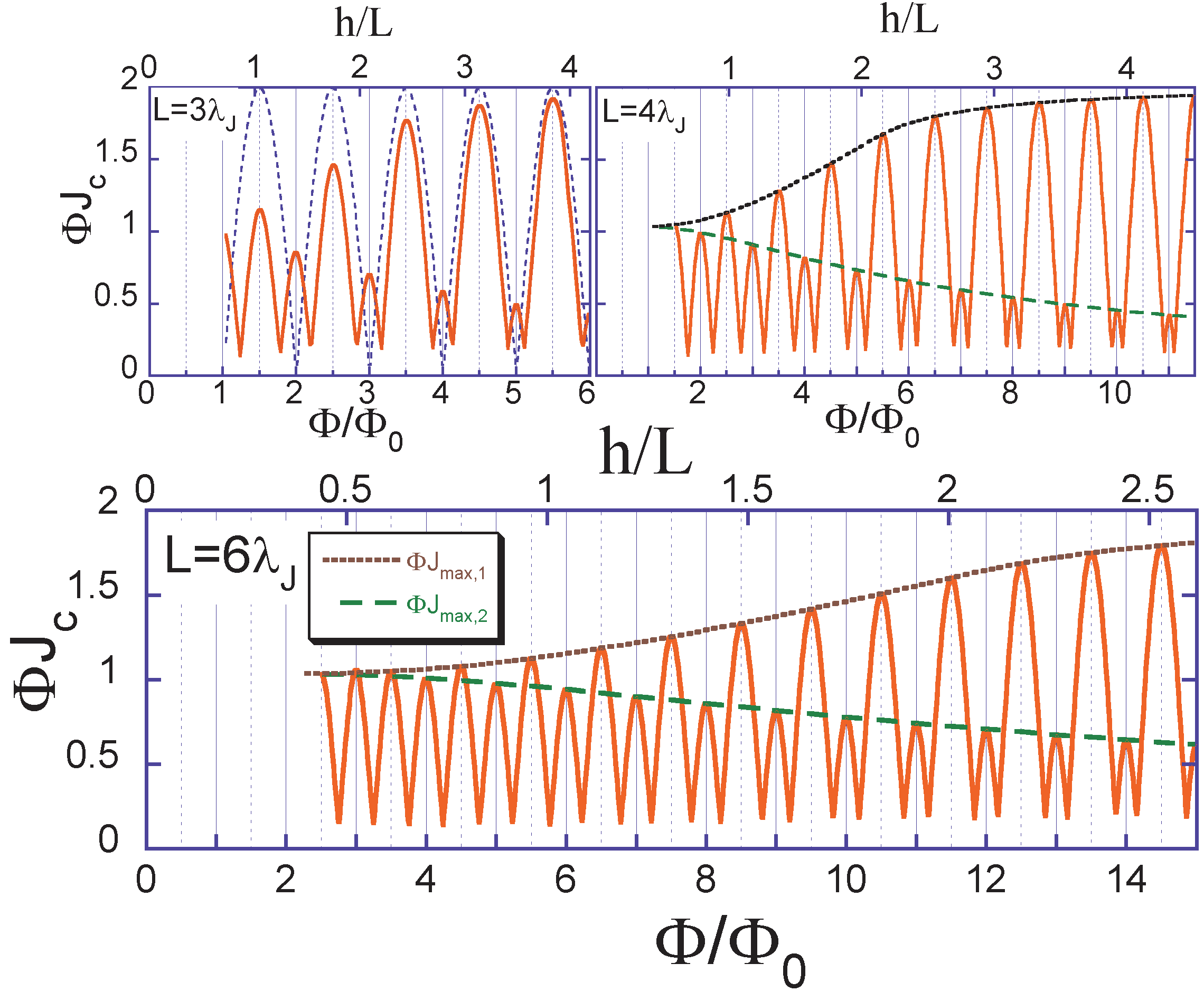}
\end{center}
\caption{The field dependences of the critical current for three different
sizes $L=3$, $4$, and $6$. To emphasize the periodic nature of dependences, we
plot the product $\Phi J_{c}$ (in units of $J_{J}\Phi_{0}/2\pi$ where $J_{J}$
is the total maximum Josephson current) vs $\Phi/\Phi_{0}$. One can observe
the crossover from $\Phi_{0}/2$-periodic oscillations to $\Phi_{0}$-periodic
oscillations with increasing $\Phi$. At larger $L$ the crossover takes place
at larger $\Phi/\Phi_{0}$. The dashed line in the $L=3\lambda_{J}$ plot shows
function $2|\sin(\pi\Phi/\Phi_{0})|$ corresponding to usual Fraunhofer
dependence in small Josephson junctions. Top axes show the parameter $h/L$ (in
real units $h/L=B/B_{L}$ with $B_{L}=L\Phi_{0}/(2\pi\gamma^{2}s^{3})$). The
crossover always takes place at the same value of ratio $h/L\sim1.5$. In the
plots for $L=4\lambda_{J}$ and $6\lambda_{J}$ dotted and dashed lines show
universal dependences of product $\Phi$ times current maxima at $\Phi
=(k+1/2)\Phi_{0}$ ($\Phi J_{max,1}$) and $\Phi=k\Phi_{0}$ ($\Phi J_{max,2}$)
on the parameter $h/L$. }
\label{Fig-Jc-h}
\end{figure}\begin{figure}[ptb]
\begin{center}
\includegraphics[width=3.in]{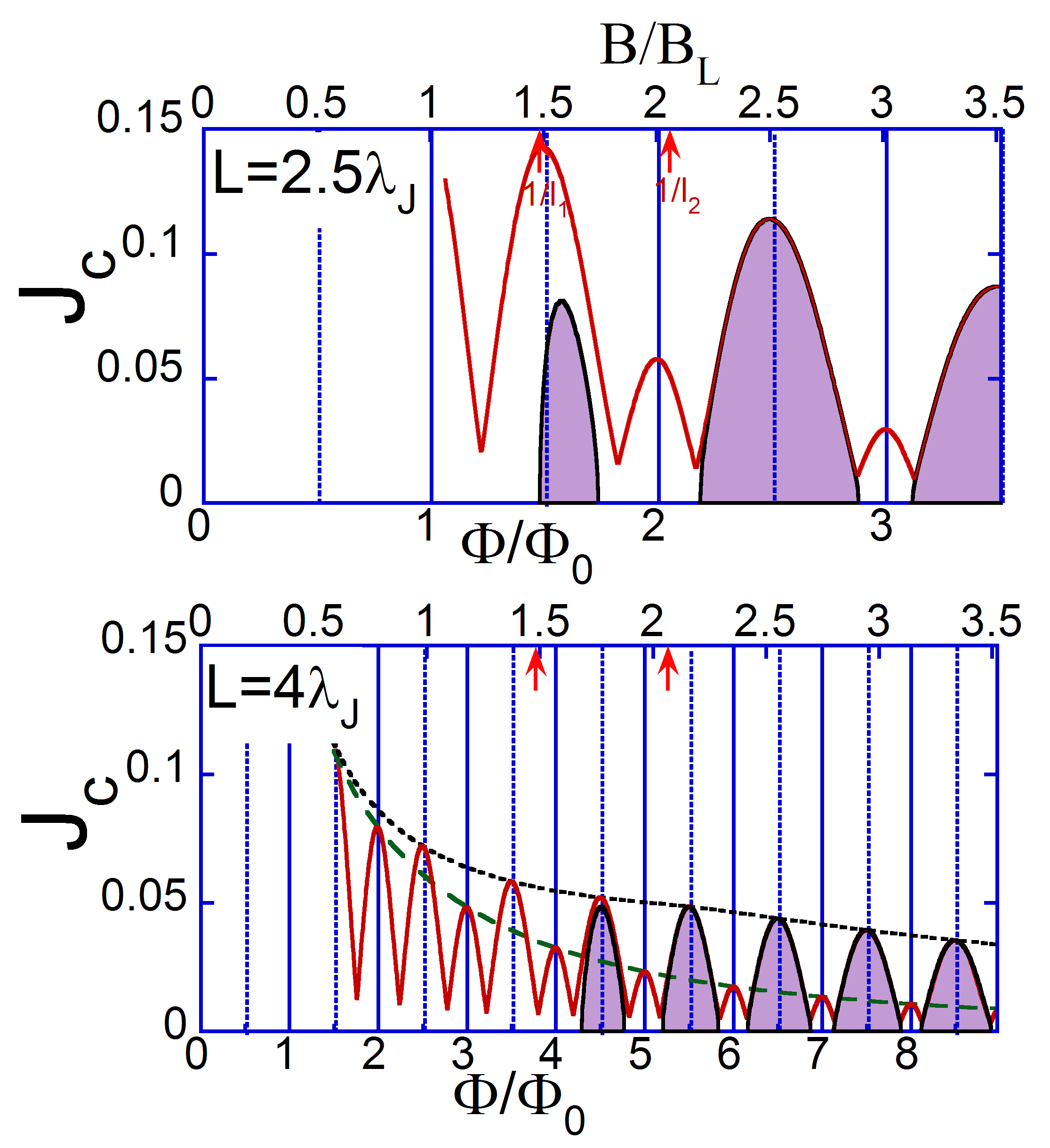}
\end{center}
\caption{The field dependences of the critical current (solid lines) for sizes
$L=2.5$, and $4$ for the same range of the ratio $h/L$, $h/L\lesssim3.5 $
shown on the top axes. Shaded areas show the regions of stable rectangular
lattice. The rectangular-lattice regions first appear in the vicinity of
points $\Phi=(k+1/2)\Phi_{0}$ when $h/L$ exceeds $1/l_{1}\approx1.48$. When
$h/L$ exceeds $1/l_{2}=2.07$ the rectangular lattice remains stable at these
points up to the critical current.}
\label{Fig-Jc-hRect}
\end{figure}

In the whole range of fields and sizes we explore the phase diagram
numerically. To find the ground state and the critical current at
given $h$ and $L$, we study dependences of the lattice structure,
energy, and Josephson current on the lattice phase shift $\alpha$.
First, we have to find the boundary deformations $\varphi_{0} $,
$\varphi_{L}$ and the elliptic parameter $m$ using the boundary
conditions (\ref{sin_phi_mon}) together with either Eq.\
(\ref{Eq_m_monot}) for $\cos\alpha\cos(hL\!-\!\alpha)\!>\!0$ or Eq.\
(\ref{Eq_m_nonmon}) for $\cos\alpha \cos(hL\!-\!\alpha)\!<\!0$.
Using obtained values, we compute the energy from Eq.\
(\ref{EnMonot}) or (\ref{EnNonMon}) and the current from Eq.\
(\ref{CurrMonot}). This procedure has been implemented in
Mathematica. Figure \ref{Fig-L4alpha} shows representative
$\alpha$-dependences of the energy and current for $L=4$ and three
values of flux per junction $\Phi/\Phi _{0}\!=\!3.1$, $3.25$, and
$3.4$ within one oscillation period. The minimum of the energy with
respect to $\alpha$ determines the ground state and the maximum of
the current determines the critical current.

Figure \ref{Fig-Jc-h} shows the field dependences of the critical
current for three different values of the junction size $L$, $3$,
$4$, and $6$. One can observe that with increasing field
$\Phi_{0}/2$-periodic oscillations smoothly transform into
$\Phi_{0}$-periodic oscillations. This occurs via suppression of the
peaks at $\Phi=k\Phi_{0}$ and enhancement of the peaks at
$\Phi=(k+1/2)\Phi_{0}$. Such behavior of the critical current has
been recently reproduced by numerical simulations by Irie and
Oya.\cite{IrieOyaSST07} Experimentally, the crossover between
$\Phi_{0}/2$- and $\Phi_{0}$-periodic oscillations has been observed
in the flux-flow voltage by Kakeya \textit{et al.}\cite{Kakeya05}
The crossover field can be arbitrarily defined as a field at which a
$k\Phi_{0}$-peak in the product $\Phi J_{c}$ drops below the half of
a $(k+1/2)\Phi_{0}$-peak. At larger $L$ the crossover takes place at
larger field and larger $\Phi /\Phi_{0}$ but it always occurs at the
same ratio $h/L$, $h/L\sim1.6$. Important property of the system,
discussed in Sec.\ \ref{Sec:Rect}, is the transformation of the
lattice into the rectangular state at sufficiently large $h/L$. This
property is illustrated in Fig.\ \ref{Fig-Jc-hRect} where behavior
of the critical current is shown together with regions of stable
rectangular lattice in the field-current diagram for two sizes,
$L=2.5$ and $4$.

Lets consider in more detail behavior of the critical-current maxima
at $\Phi=(j/2)\Phi_{0}$. We start with the case of half-integer flux
quanta per junction, $\Phi=(k+1/2)\Phi_{0} $\ ($hL=\left(
2k+1\right) \pi$). In this case a nonmonotonic solution is realized
for all $\alpha$. As for sufficiently high $h/L$ the lattice
transforms into rectangular state, it will be convenient to use
presentation given by Eqs.\ (\ref{mon_psi0}),\ (\ref{mon_psiL}), and
(\ref{nonmon_psi_m}) which naturally describes this transition. As
follows from Eq.~(\ref{Condit-Rect}), in the range
\begin{equation}
\cos \alpha  >\tan\left(  \sqrt{2}L/h\right)  /\sqrt{2}
\label{Cond-Rect-half-int}
\end{equation}
the rectangular lattice is realized for which $\tilde{m}=0$ and the energy and
current are given by
\begin{align*}
\mathcal{E}_{1}(\alpha)  &  =-\frac{2}{h}\left\vert \cos \alpha
\right\vert \\
J_{1}(\alpha)  &  =\frac{2}{h}\sin \alpha\ \mathrm{sign}\!\left[
\cos \alpha \right]
\end{align*}
In the opposite range, $\cos \alpha\!<\!\tan\left(  \sqrt
{2}L/h\right)  $$/$$\sqrt{2}$, the solution has the form of deformed
lattice. In this case we have $\psi_{0}\!=\!\psi_{L}\!=\!\arcsin(
1/\sqrt{1\!+\!2\cos^{2} \alpha})  $ (or $\pi\!-\!\arcsin(
1/\sqrt{1\!+\!2\cos^{2} \alpha})$ ) and Eqs.\ (\ref{nonmon_psi_m})
becomes
\begin{equation}
K\left(  \tilde{m}\right)  \!-\!F\left(  \arcsin\left(  \frac{1}
{\sqrt{1\!+\!2\cos^{2} \alpha}}\right)  ,\tilde{m}\right)
\!=\!\frac{\sqrt
{2}L}{h}. \label{half-int-m}
\end{equation}
Solving this equation with respect to the elliptic parameter $\tilde{m}
(\alpha)$, we can obtain the energy and current
\begin{align}
\mathcal{E}_{1}(\alpha)  &  =-\frac{2}{h}\left\vert \sqrt{1-\frac{\tilde{m}
}{1+2\cos^{2}\left(  \alpha\right)  }}\cos\left(  \alpha\right)  \right\vert
\nonumber\\
+  &  \frac{\sqrt{2}}{h}\left[  E(\tilde{m})-E(\psi_{0},\tilde{m})\right]
-\frac{\left(  2-m\right)  L}{h^{2}}\label{En-half-int}\\
J_{1}(\alpha)  &  =\frac{2}{h}\sqrt{1\!-\!\frac{\tilde{m}}{1\!+\!2\cos
^{2}\left(  \alpha\right)  }}\sin\left(  \alpha\right)  \mathrm{sign}\!\left[
\cos( \alpha) \right]  \label{Curr-half-int}
\end{align}
The critical current at $\Phi\!=\!(k+\!1/2)\Phi_{0}$ is given by
$J_{\max,1} =\max_{\alpha}[J_{1}(\alpha)]$. At small $L$, $L/h\ll1$,
the maximum critical current is realized at the instability point of
the rectangular lattice $\cos\left( \alpha\right)  \approx L/h$
giving
\begin{equation}
J_{\max,1}\approx\frac{2}{h}\left(  1-\frac{L^{2}}{2h^{2}}\right)  .
\label{CritCurr-half-small}
\end{equation}
It is always somewhat smaller than the ``Fraunhofer'' value $2/h$.

It was obtained in Section \ref{Sec:Rect} that the rectangular
lattice is realized in ground state ($\alpha=0$) at points
$\Phi=(k+1/2)\Phi_{0}$ for $L/h<l_{1}=0.675$. If, however, $L/h$ is
only slightly smaller than this value, the rectangular lattice
becomes unstable with increasing current and the configuration at
the critical current still corresponds to the deformed lattice. We
found that there is another typical value of the ratio $L/h$,
$L/h=l_{2}\approx0.484$, below which \emph{the rectangular lattice
remains stable up to the critical current}. Both typical values of
$h/L$, $1/l_{1}$ and $1/l_{2}$, are marked in Fig.\
\ref{Fig-Jc-hRect}. One can see in that for both shown stack sizes,
$L=2.5$ and $4$, the rectangular lattice first appears around points
$\Phi=(k+1/2)\Phi_{0}$ when $h/L$ exceeds $1/l_{1}$ and its
stability range extends up to the critical current when $h/L$
exceeds $1/l_{2}$.

For integer flux quanta, $\Phi\!=\!k\Phi_{0}$ ($hL\!=\!2\pi k$), a
changing-sign monotonic solution always realizes,
$v_{L}\!=\!-v_{0}$. In the case $\cos\alpha\!>\!0$ this corresponds
to $\varphi_{0}=\pi-\varphi_{L}=\arcsin[m(1+2\cos ^{2}\left(
\alpha\right)  )]^{-1/2}$ and Eq.\ (\ref{Eq_m_monot}) can be reduced
to the form
\begin{equation}
\sqrt{m}\left(  K\left(  m\right)  -F\left(  \phi_{0},m\right)  \right)
=\sqrt{2}L/h. \label{int-m}
\end{equation}
Solving this equation with respect to $m$, we can obtain the energy from
Eq.\ (\ref{EnMonot}) and current from Eq.\ (\ref{CurrMonot})
\begin{align}
&  \mathcal{E}_{2} \! =\!-\frac{2}{h}\cos\phi_{0}\cos \alpha
\!+\!\frac{\sqrt{2}}{\sqrt{m}h}\left[ E(m)\!-\!E(\phi_{0},m)\right]
\!
-\!\frac{L}{mh^{2}}\label{En-int}\\
&  J_{2}(\alpha) =\frac{2}{h}\cos\phi_{0}\sin\alpha\label{Curr-int}
\end{align}
The critical current at $\Phi\!=\!(k\!+\!1/2)\Phi_{0}$ is given by
$J_{\max,2}\! =\!\max_{\alpha}[J_{2}(\alpha)]$. At small $L$
inequality $\cos\phi_{0}\ll1$ holds. In this limit, using relation
$F(\phi_{0},m)\approx K\left(  m\right) \!-\!\left(
\pi/2\!-\!\phi_{0}\right)  /\sqrt{1\!-\!m}$, we can approximately
rewrite Eq.\ (\ref{int-m}) as
$\pi/2\!-\!\phi_{0}=(\sqrt{2}L/h)\sqrt{1\!-\!1/m}$. As $\sin
\phi_{0}$ is close to one, Eq.\ (\ref{sin_phi0_mon}) gives
$1/m\approx 1\!+\!2\cos^{2}\left(  \alpha\right)  $ and
$\phi_{0}\!=\!\pi/2\!-\!(2L/h)\cos\alpha$. Therefore we obtain for
the $\alpha$-dependent current (\ref{Curr-int}),
\begin{equation}
J_{2}(\alpha)\approx\frac{2L}{h^{2}}\sin2\alpha. \label{Curr-alpha-int-small}
\end{equation}
The maximum is realized at $\alpha=\pi/4$ giving the following result for the
critical current
\begin{equation}
J_{\max,2}\approx 2L/h^{2}, \label{CritCurr-int-small}
\end{equation}
i.e., it decays at large $h$ as $1/h^{2}$ but never drops to zero as for usual
Fraunhofer dependence. The behavior in the narrow-stack regime will be
considered in more details in the next section. One can see that the critical
currents at both maxima $J_{\max,\alpha}$ ($\alpha=1,2$) have the same scaling
property: the product $hJ_{\max,\alpha}$ depends only on the ratio $L/h$.
These scaling dependences are plotted in Fig.\ \ref{Fig-Jc-h} in the plots for
$L=4$ and $6$.

\section{Narrow-stack regime \label{Sec:NarrowStack}}

Lets consider in more detail the narrow-stack regime at $L/h\ll1$. In this
regime interaction with the boundaries is typically stronger than the bulk
shearing interaction. As a consequence, the boundaries stabilize the
rectangular lattice configuration in most part of the phase diagram. The
exception is the narrow regions in the vicinity of the integer-flux-quanta
points $\Phi=k\Phi_{0}$ where the interaction with the boundaries vanishes and
the rectangular lattice looses its stability. The rectangular lattice also
becomes unstable near the critical current. In this section we will study in
details this behavior. Instead of using asymptotic behavior of elliptic
integrals, it is more transparent to use as a starting point the equation for
smooth alternating deformation (\ref{EqSmooth}), the boundary conditions
(\ref{BoundSmooth}) and the energy (\ref{EnSmooth}). It will be more
convenient to use variable $\varphi$ given by Eq.\ (\ref{ElliptVar}) (instead
of $v$) from the very beginning, because it vanishes in the
rectangular-lattice state. We also introduce a new variable for the lattice
phase shift,
\[
\beta\equiv\alpha+\pi/2-hL/2,
\]
which will facilitate a more compact presentation of results. In terms of the
variables $\varphi(u)$ and $\beta$ the energy (\ref{EnSmooth}) can be
rewritten as
\begin{align}
\mathcal{E}(\beta)  &  \!\approx\!-\frac{1}{h}\left[  \cos\varphi_{0}
\sin\!\left(  \beta\!+\!\frac{hL}{2}\right)  \! -\!\cos\varphi_{L}
\sin\!\left(  \beta\!-\!\frac{hL}{2}\right)  \right] \nonumber\\
&  +\int_{0}^{L}du\left[  \frac{1}{8}\left(  \frac{d\varphi}{du}\right)
^{2}-\frac{1-\cos(2\varphi)}{2h^{2}}\right]  . \label{En-phi}
\end{align}
From this energy we obtain equation for $\varphi(u)$,
\begin{equation}
\frac{d^{2}\varphi}{du^{2}}+\frac{4}{h^{2}}\sin\left(  2\varphi\right)  =0,
\label{EqSmooth-phi}
\end{equation}
and the boundary conditions
\begin{subequations}
\label{BoundCond-phi}
\begin{align}
\frac{d\varphi}{du}(0)  &  =\frac{4}{h}\sin\left(  \varphi_{0}\right)
\sin\left(  \beta+\frac{hL}{2}\right)  ,\ \label{BoundCond-phi0}\\
\frac{d\varphi}{du}(L)  &  =\frac{4}{h}\sin\left(  \varphi_{L}\right)
\sin\left(  \beta-\frac{hL}{2}\right)  . \label{BoundCond-phiL}
\end{align}

For small $L$ Eq.\ (\ref{EqSmooth-phi}) can be solved as expansion with
respect to powers of $u-L/2$,
\end{subequations}
\begin{equation}
\varphi=\varphi_{a}+a\left(  u-\frac{L}{2}\right)  -\frac{2\left(
u-L/2\right)  ^{2}}{h^{2} }\sin\left(  2\varphi_{a}\right)
\label{PhaseExpanSmall}
\end{equation}
with $\varphi_{a}=\varphi(L/2)$. Boundary conditions (\ref{BoundCond-phi})
give two equations for two unknown variables, the midpoint phase $\varphi_{a}
$ and the linear slope $a$. We obtain two types of solutions: (i) the
rectangular-lattice solution $a=0$, $\sin\varphi_{a}=0$ and (ii) the
deformed-lattice solution. In the leading order with respect to the small
parameter $L/h$, the latter solution can be represented as
\begin{align}
a  &  \approx\frac{4}{h}\sin\left(  \varphi_{a}\right)  \sin\beta\cos\left(
hL/2\right)  ,\label{Slope}\\
\cos\left(  \varphi_{a}\right)   &  \approx\frac{h}{L}\frac{\sin( hL/2)
\cos\beta}{1+2\sin^{2}\!\beta\cos^{2}(hL/2)}. \label{phi-av}
\end{align}
As follows from the last equation, the deformed-lattice solution does not
exist if
\begin{equation}
\frac{h}{L}\frac{\left\vert \sin\left(  hL/2\right)  \cos\beta\right\vert
}{1+2\sin^{2}\beta\cos^{2}\left(  hL/2\right)  }>1. \label{Condition}
\end{equation}
In this case the configuration must be the rectangular lattice. The solution
(\ref{phi-av}) also includes the case of the ideal triangular lattice
$\varphi_{a} =\pi/2$ which is always realized if either $\sin\left(
hL/2\right)  =0$ or $\cos\beta=0$. As we consider the region $L/h\ll1$, both
triangular and deformed lattices exist only in vicinity of these points.

\begin{figure}[ptb]
\begin{center}
\includegraphics[width=3.2in]{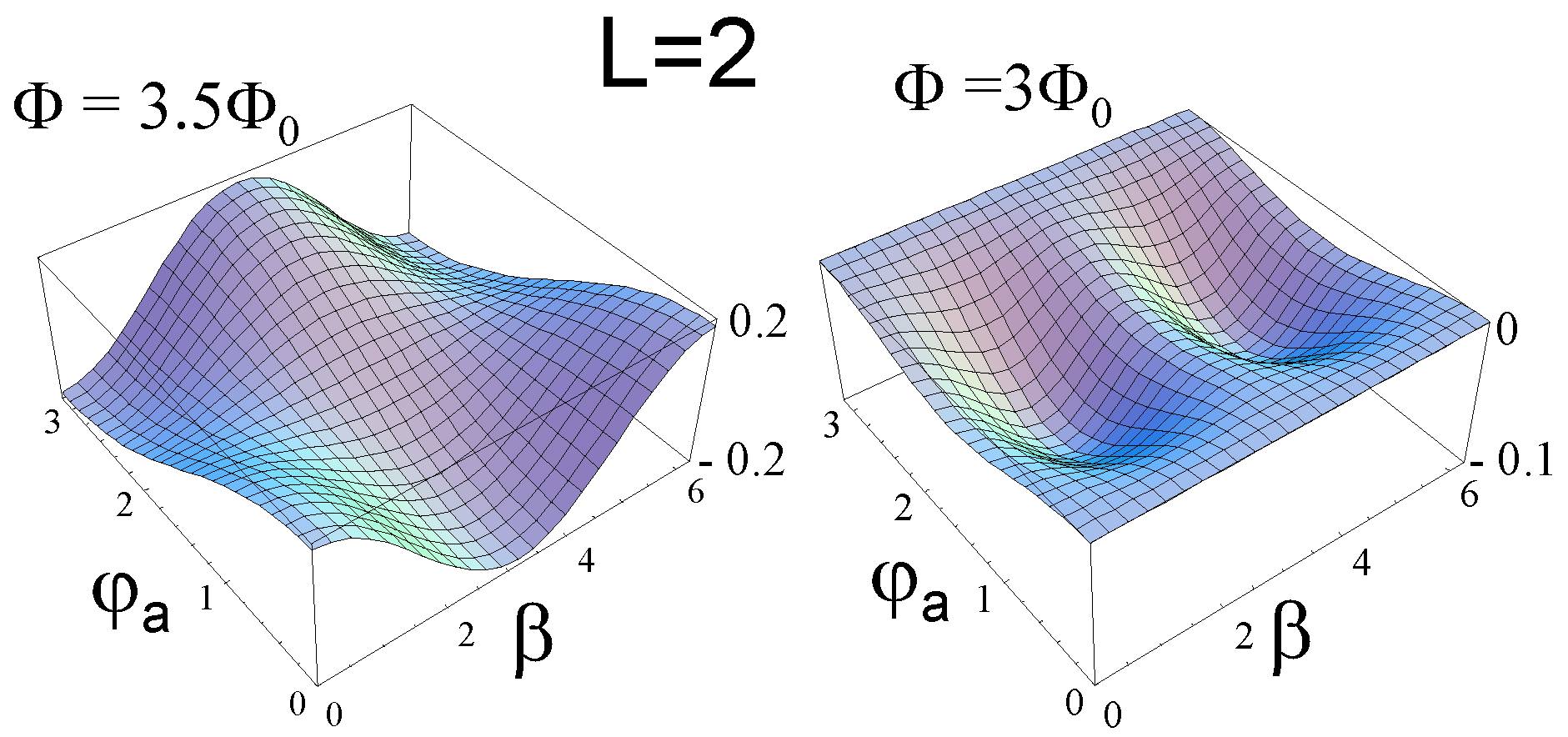}
\end{center}
\caption{Examples of the energy landscape (\ref{En-phi-beta}) as a function of
the average lattice shift $\beta$ and the amplitude of the alternating phase
deformation $\phi_{a}$ for $L=2$ and $\Phi/\Phi_{0}=$ 3.5 and 3. For $\Phi=
3.5 \Phi_{0}$ the equivalent minima at $(\phi_{a},\beta)=(0,\pi)$ and
$(\pi,0)$ correspond to the rectangular lattice, while for $\Phi= 3\Phi_{0}$
the minima at $(\phi_{a},\beta)=(\pi/2,\pi/2)$ and $(\pi/2,3\pi/2)$ correspond
to the triangular lattice.}
\label{Fig-En-profile}
\end{figure}

For analysis of lattices, it is also useful to derive the energy as a function
of the average lattice shift, $\beta$, and the relative phase shift between
the neighboring layers, $\phi_{a}$. For that we substitute expansion
(\ref{PhaseExpanSmall}) up to the linear order with the phase gradient given
by Eq.\ (\ref{Slope}) into the energy (\ref{En-phi}) and obtain
\begin{align}
&  \mathcal{E}(\varphi_{a},\beta)\approx-\frac{2}{h}\cos\left(  \varphi
_{a}\right)  \sin\left(  \frac{hL}{2}\right)  \cos\beta\nonumber\\
&  -\frac{L}{h^{2}} \sin^{2}\left(  \varphi_{a}\right)  \left[  1+2\sin
^{2}\beta\cos^{2}\left(  \frac{hL}{2}\right)  \right]  \label{En-phi-beta}
\end{align}
In particular, the result (\ref{phi-av}) corresponds to the minimum
of this energy with respect to $\varphi_{a}$ when the condition
(\ref{Condition}) is satisfied. We will see that this relatively
simple energy function of two variables, whose shape evolves with
the magnetic field, describes a surprisingly rich behavior in the
vicinity of the integer-flux-quanta points. The typical energy
landscapes for the cases of half-integer and integer flux quanta per
junction are illustrated in Fig.\ \ref{Fig-En-profile}.

Lets study zero-current ground states first. We start with stability
analysis for the rectangular lattice which is realized at
$\varphi_{a}\!=\!0,\pi$ and gives ground state in the most part of
phase space. In this case the energy is minimal with respect to
$\beta$ either at $\beta=0$, for $\cos\left( \varphi_{a}\right)
\sin\left(  hL/2\right)  >0$, or at $\beta=\pi$, for $\cos\left(
\varphi_{a}\right)  \sin\left(  hL/2\right)  <0$, see, e.g., left
part of Fig.\ \ref{Fig-En-profile} where $\sin\left(  hL/2\right)
=-1$. Expanding the energy with respect to $\varphi_{a}$ near the
point $\varphi _{a}=0$,
\[
\mathcal{E}(\varphi_{a},0)\!\approx\!-\frac{2}{h}\left\vert \sin\left(
\frac{hL}{2}\right)  \right\vert +\frac{\varphi_{a}^{2}}{h}\left(  \left\vert
\sin\left(  \frac{hL}{2}\right)  \right\vert -\frac{L}{h}\right)  ,
\]
we conclude that the rectangular lattice is stable if
\begin{equation}
\left\vert \sin\left(  \frac{hL}{2}\right)  \right\vert >\frac{L}{h},
\label{Stab-rect}
\end{equation}
which coincides with the condition (\ref{Condition}) at $\beta=0,\pi$. In real
variables the condition (\ref{Stab-rect}) can be rewritten as
\begin{equation}
\frac{2\pi\Phi}{\Phi_{0}}\left\vert \sin\left(  \frac{\pi\Phi}{\Phi_{0}
}\right)  \right\vert >\left(  \frac{L}{\lambda_{J}}\right)  ^{2}.
\label{Stab-rect-real}
\end{equation}
Therefore even at small $L/h$ the rectangular lattice is always
unstable near the integer-flux-quanta points, $\Phi=k\Phi_{0}$. This
is easy to understand: near these points the interaction with the
boundaries vanishes and even small shearing interaction between the
neighboring planar Josephson-vortex arrays becomes sufficient to
induce instability with respect to the alternating deformations.
Further analysis, however, will show that this instability takes
place when the rectangular lattice does not give already the ground
state, meaning that the system actually experiences a first-order
transition.

For the deformed-lattice solution (\ref{phi-av}) the energy at fixed $\beta$
is given by
\begin{equation}
\mathcal{E}(\beta)\!\approx\!-\!\frac{1}{L}\frac{\sin^{2} \frac{hL} {2}
\cos^{2}\beta}{1\!+\!2\sin^{2}\!\beta\cos^{2}\!\frac{hL}{2}}\!-\!\frac
{L}{h^{2}}\left(  1\!+\!2\sin^{2}\!\beta\cos^{2}\!\frac{hL}{2}\right)  .
\label{En-def-lat-beta}
\end{equation}
This energy always has extremums at $\beta=0$, $\pi$, and $\pi/2$. As follows
from Eq.\ (\ref{phi-av}), the state at $\beta=\pi/2$ always corresponds to the
triangular lattice, $\varphi_{a}=\pi/2$. We find now the conditions that the
energy reaches minima at these values of $\beta$. Consider first the point
$\beta=0$. Expanding the energy near this point,
\begin{align*}
\mathcal{E}(\beta)  &  \approx-\frac{1}{L}\sin^{2}\!\frac{hL}{2}
-\frac{L}{h^{2}}\\
&  +\frac{\beta^{2}}{L}\left[  \sin^{2}\!\frac{hL}{2} \left(
1\!+\!2\cos^{2}\frac{hL}{2}\right)
\!-\!\frac{2L^{2}}{h^{2}}\cos^{2}\frac{hL} {2}\right]  ,
\end{align*}
we obtain that it corresponds to minimum if
\[
\tan^{2}\left(  \frac{hL}{2}\right)  \left[
1+2\cos^{2}\left(\frac{hL}{2}\right)\right]
>\frac{2L^{2}}{h^{2}}.
\]
As $L/h\ll1$, this inequality is valid almost everywhere except in
the vicinity of the integer-flux-quanta points at which
$\sin(hL/2)\rightarrow0$ and $\cos^{2}(hL/2)\rightarrow1$ where this
condition can be rewritten in an approximate simpler form,
\[
\left\vert \sin\left(  \frac{hL}{2}\right)  \right\vert >\frac{\sqrt{2}
L}{\sqrt{3}h}.
\]
Comparing this condition with the condition (\ref{Stab-rect}), we can see that
the deformed lattice gives the energy minimum at $\beta=0$ only within the
narrow region given by
\begin{equation}
\sqrt{\frac{2}{3}}<\frac{h}{L}\left\vert \sin\left(  \frac{hL}{2}\right)
\right\vert <1.
\end{equation}
In this region the optimum value of $\varphi_{a}$ for $\beta=0$ is given by
$\cos\left(  \varphi_{a}\right)  =(h/L)\left\vert \sin\left(  hL/2\right)
\right\vert $.

To find if the triangular lattice at the point $\beta=\pi/2$ gives the local
energy minimum, we expand the energy (\ref{En-def-lat-beta}) near this point,
$\beta=\pi/2-\zeta$,
\begin{align*}
\mathcal{E}(\beta)  &  \approx-\frac{L}{h^{2}}\left(  1+2\cos^{2}\frac{hL}
{2}\right) \\
&  +\zeta^{2}\left(  -\frac{1}{L}\frac{\sin^{2}\frac{hL}{2} }{1\!+\!2\cos
^{2}\!\frac{hL}{2}}+\frac{2L}{h^{2}}\cos^{2}\frac{hL}{2}\right)  .
\end{align*}
We can see that the value $\beta=\pi/2$ corresponds to energy minimum if
\[
\frac{\tan^{2} \frac{hL}{2}}{1\!+\!2\cos^{2}\!\frac{hL}{2}}<\frac{2L^{2}
}{h^{2}}
\]
or, approximately, $|\sin(hL/2)|<\sqrt{6}L/h$. We can conclude that near the
integer-quanta points $\Phi=k\Phi_{0}$ ($hL=2k\pi$) the minimum location
switches from $\beta=0$ to $\beta=\pi/2$. In the intermediate region given
approximately
\begin{equation}
\sqrt{\frac{2}{3}}<\frac{h}{L}\left\vert \sin\left(  \frac{hL}{2}\right)
\right\vert <\sqrt{6} \label{Two-minima}
\end{equation}
the energy has local minimums at both points, $\beta=0$ and $\pi/2$. Moreover,
in the region $(h/L)\left\vert \sin(hL/2) \right\vert >1$ the minimum at
$\beta=0$ is realized by the rectangular lattice. This behavior indicates that
\emph{switching between the rectangular and triangular lattices in the ground
state occurs via a first-order phase transition}.

\begin{figure}[ptb]
\begin{center}
\includegraphics[width=3.2in]{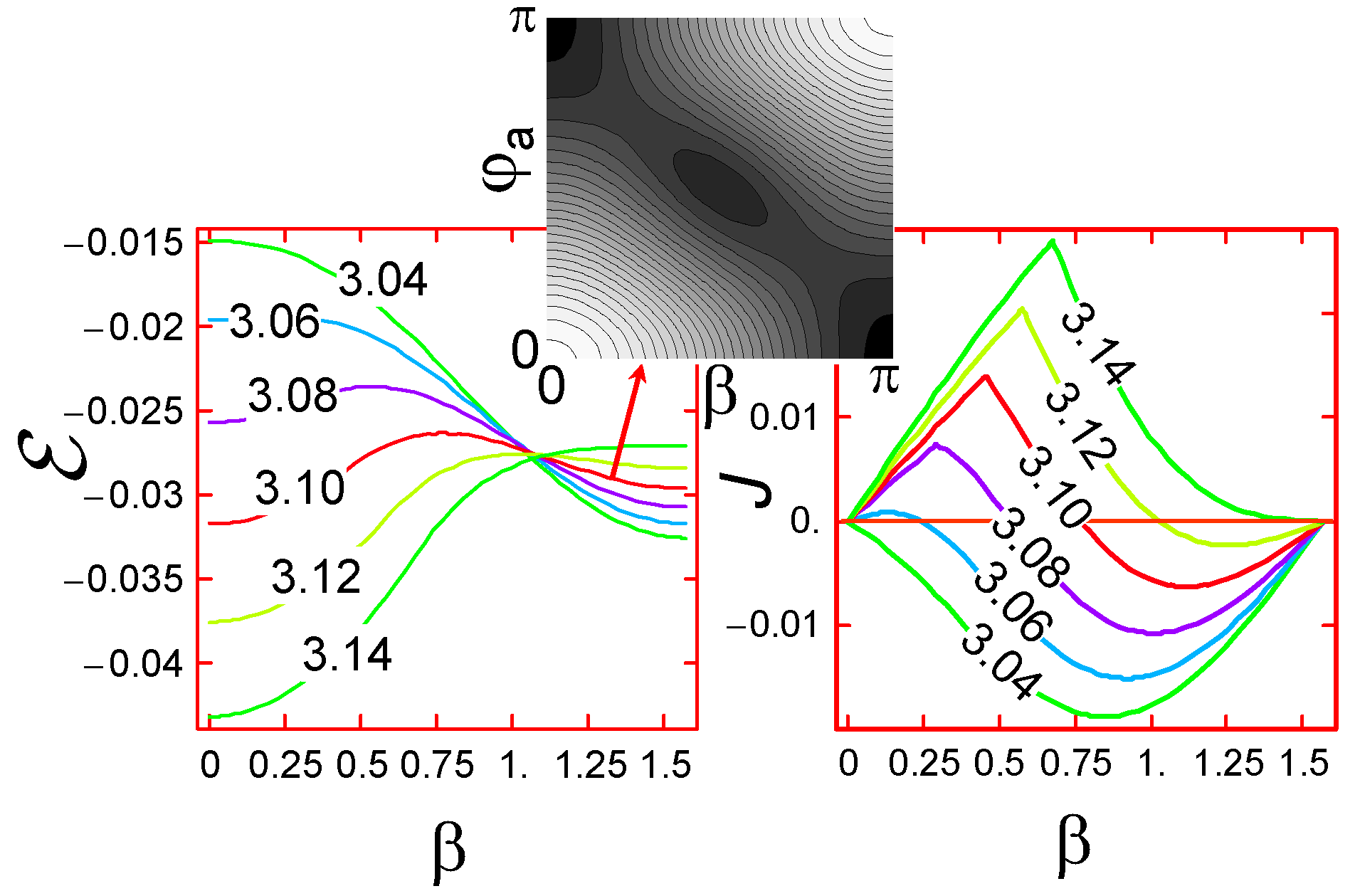}
\end{center}
\caption{The dependences of the energy (left) and current (right) on the
lattice phase shift $\beta$ for $L=2\lambda_{J}$ slightly above the point
$\Phi=3\Phi_{0}$ (curves are marked by the values of $\Phi/\Phi_{0}$). One can
observe several features discussed in details in the text. Close the
$\Phi=3\Phi_{0}$ for $\Phi<3.1\Phi_{0}$ the global energy has minimum at
$\beta=\pi/2$ corresponding to the triangular lattice (Eq.\ (\ref{f-tk})
actually gives $\Phi_{t}=(3+1/\pi^{2})\Phi_{0}$ for the transition point).
Above this point the global minimum is given by the rectangular lattice at
$\beta=0$. Within narrow range of $\Phi$ the energy has minima at both
$\beta=0$ and $\pi/2$. The rectangular lattice appears at the local minimum at
$\beta=0$ for $\Phi\geq3.08\Phi_{0}$ but it becomes unstable with increasing
$\beta$. The instability points are seen as kinks in the $J(\beta)$ curves. In
some range of $\Phi$ the absolute value of current has two local maxima within
$0<\beta<\pi/2$. The critical current switches between these maxima with
increasing $\Phi$. Inset shows the contour plot of energy near the transition
point, at $\Phi=3.1\Phi_{0}$.}
\label{Fig-En-J-beta}
\end{figure}

To find the transition point, we compare the triangular-lattice energy,
\[
\mathcal{E}_{\mathrm{trian}}\equiv\mathcal{E}\left(  \frac{\pi}{2},\frac{\pi}
{2}\right)  =-\frac{L}{h^{2}}\left(  1+2\cos^{2}\frac{hL}{2}\right)  ,
\]
with the rectangular-lattice energy,
\[
\mathcal{E}_{\mathrm{rect}}\equiv\mathcal{E}(0,0)=-\frac{2}{h}\left\vert
\sin\left(  \frac{hL}{2}\right)  \right\vert ,
\]
and obtain that the triangular lattice wins if
\[
\left\vert \sin\left(  \frac{hL}{2}\right)  \right\vert <\frac{L}{2h}\left[
1+2\cos^{2}\left(  \frac{hL}{2}\right)  \right]  .
\]
As this only happens near the points where $\cos^{2}(hL/2)\approx1$, the
equation for the transition points, $h_{t}$, can again be rewritten in a
simpler form,
\begin{equation}
\left\vert \sin\left(  \frac{h_{t}L}{2}\right)  \right\vert =\frac{3}{2}
\frac{L}{h_{t}}, \label{first-order-tran}
\end{equation}
or in real units, in terms of flux per junction,
\begin{equation}
\frac{2\pi\Phi_{t}}{\Phi_{0}}\left\vert \sin\left(  \frac{\pi\Phi_{t}}
{\Phi_{0}}\right)  \right\vert =\frac{3}{2}\left(  \frac{L}{\lambda_{J}
}\right)  ^{2}. \label{first-order-tran-real}
\end{equation}
Comparing Eq.\ (\ref{first-order-tran}) with the stability criterion of the
rectangular lattice (\ref{Stab-rect}), we indeed can see that before the
rectangular lattice becomes unstable, it switches to the triangular lattice
via a first-order phase transition. From this equation we can also obtain
small shift of transition point with respect to the integer-flux-quanta point
$\Phi=k\Phi_{0}$ as a function of the index $k$. Writing $\Phi_{t}=\left(
k+f_{t,k}\right)  \Phi_{0}$, we compute
\begin{equation}
\left\vert f_{t,k}\right\vert \approx\frac{3}{4\pi^{2}k}\left(  \frac
{L}{\lambda_{J}}\right)  ^{2}\ll1. \label{f-tk}
\end{equation}
The discussed behavior is illustrated in Fig.\ \ref{Fig-En-J-beta} in which
the $\beta$-dependences of the energy and current are plotted for $L=2$ and
several values of $\Phi$ above the point $3\Phi_{0}$. The contour plot of
energy at the transition point is also shown. 

Let us investigate now current-carrying states and behavior of the critical
current. In the region given by Eq.\ (\ref{Condition}) the current in the
rectangular-lattice state is
\begin{equation}
J(\beta)=\frac{2}{h}\left\vert \sin\left(  \frac{hL}{2}\right)  \right\vert
\sin\beta, \label{Curr-beta-rect}
\end{equation}
where we assumed for definiteness that $\cos\varphi_{a} \sin(hL/2)>0$. If the
condition (\ref{Condition}) is violated then the deformed lattice is realized.
In this case, differentiating the energy (\ref{En-def-lat-beta}) with respect
to $\beta$, we obtain the current in the deformed-lattice state (including the
triangular lattice)
\begin{align}
J(\beta)  &  \!=\!\left(  \frac{\sin^{2} \frac{hL}{2} \left(
1+2\cos^{2} \frac{hL}{2}\right)  }{L\left(
1+2\sin^{2}\beta\cos^{2}\frac{hL}{2}\right)
^{2}}\!-\!\frac{2L\cos^{2}\frac{hL}{2}}{h^{2}}\right)  \! \sin2\beta
,\label{Curr-beta-def}\\
&  \text{for }\frac{h}{L}\frac{\left\vert \sin\frac{hL}{2} \cos\beta
\right\vert }{1+2\sin^{2}\beta\cos^{2}\frac{hL}{2}}<1.\nonumber
\end{align}
In particular, at $hL=2\pi k$ this formula reproduces results
(\ref{Curr-alpha-int-small}) and (\ref{CritCurr-int-small}) obtained from the
elliptic-integral representation.

Consider first the region where the ground state is given by the rectangular
lattice. Maximum current in this state would be achieved at $\beta=\pi/2$ but
the condition (\ref{Condition}) always breaks down before that. From this
condition we compute the value of $\beta$ at which the rectangular lattice
becomes unstable
\begin{equation}
|\cos\beta_{t}|\!=\!\frac{2\frac{L}{h}\left(
1\!+\!2\cos^{2}\frac{hL} {2}\right)  }{\left\vert \sin\!\frac{hL}{2}
\right\vert \!+\!\sqrt{\sin ^{2}\!\frac{hL}{2}
\!+\!\frac{8L^{2}}{h^{2}}\cos^{2}\!\frac{hL} {2}\left(
1\!+\!2\cos^{2}\!\frac{hL}{2}\right)  }}. \label{Instab-beta}
\end{equation}
In a wide range of parameters, away from the regions given by
Eq.\ (\ref{first-order-tran}), the maximum current is achieved at this
instability point
\begin{equation}
J_{c}=\frac{2}{h}\left\vert \sin\left(  \frac{hL}{2}\right)  \right\vert
\sin\beta_{t}. \label{CritCurr-Inst}
\end{equation}
In particular, in most part of the parameter space, for $|\tan(hL/2)|\gg L/h$
we obtain much simpler results
\begin{align}
&  |\cos\beta_{t}| \approx\frac{L}{h}\frac{1+2\cos^{2}\frac{hL}{2}
}{\left\vert \sin\frac{hL}{2} \right\vert },\label{Instab-beta-Appr}\\
&  J_{c} \approx\frac{2}{h}\left\vert \sin\left(  \frac{hL}{2}\right)
\right\vert \left[  1-\left(  \frac{L}{h}\right)  ^{2}\frac{\left(
1+2\cos^{2}\frac{hL}{2}\right)  ^{2}}{2\sin^{2} \frac{hL}{2} }\right]  .
\label{CritCurr-Inst-Appr}
\end{align}
In this region the critical current is only slightly smaller than the
``Fraunhofer'' result $(2/h)\left\vert \sin\left(  hL/2\right)  \right\vert $.
At the half-integer-flux-quanta points $hL=(2k+1)\pi$ these equations
reproduce result (\ref{CritCurr-half-small}) obtained from the
elliptic-integrals representation. The property that the rectangular lattice
is always unstable at some lattice displacement $\beta$ also has important
dynamic consequences. It means that the lattice can not maintain its static
rectangular configuration when it starts to move.

\begin{figure}[ptb]
\begin{center}
\includegraphics[width=3.2in]{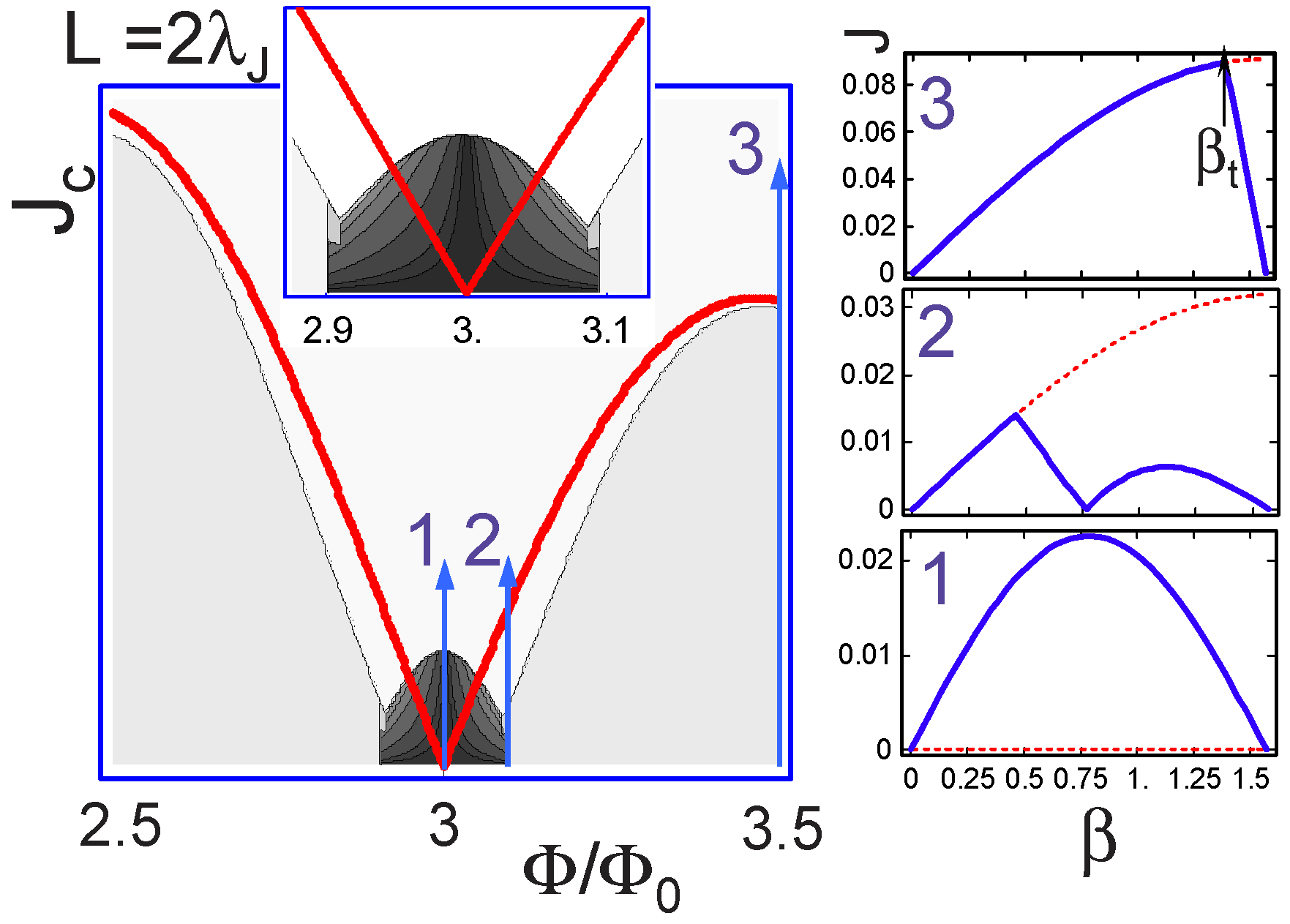}
\end{center}
\caption{The main plot: the field dependence of the critical current for
$L=2\lambda_{J}$ within one oscillation period $2.5\Phi_{0}<\Phi<3.5\Phi_{0}$.
The grey level codes the value of $|\cos\varphi_{a}|$ with the light grey
color in most part of the plot corresponding to the rectangular lattice
$|\cos\varphi_{a}|=1$ and the black color near $\Phi/\Phi_{0}=3$ corresponding
to triangular lattice $|\cos\varphi_{a}|=0$. The solid line shows the
Fraunhofer dependence $\sin(\pi\Phi/\Phi_{0})/(\pi\Phi/\Phi_{0})$. The inset
above shows blowup of the region near $\Phi/\Phi_{0}=3$. Plots at the right
side illustrate representative dependences $|J(\beta)|$ for three values of
$\Phi$ marked by arrows. Dashed curves in these plots show corresponding
dependences for usual small junction. The kinks in the $J_{c}(\Phi)$ curve at
$\Phi/\Phi_{0}\approx3\pm0.085$ occur due to the switching between different
maxima in the $|J(\beta)|$ dependence.}
\label{Fig-Jc-h-small}
\end{figure}

The critical current has a nontrivial behavior in the vicinity of points
$hL=2\pi k$ where $|\sin\left(  \frac{hL}{2}\right)  |\ll1$ and general
formula (\ref{Curr-beta-def}) can be simplified as
\[
J(\beta)\approx\frac{2L}{h^{2}}\left(  1-\frac{3h^{2}\sin^{2}\left(  \frac
{hL}{2}\right)  }{2L^{2}\left(  2-\cos2\beta\right)  ^{2}}\right)  \sin
2\beta.
\]
This gives the critical current near $hL=2\pi k$
\[
J_{c}(h,L)\approx\frac{2L}{h^{2}}\left(  1-\frac{3h^{2}\sin^{2}\left(
\frac{hL}{2}\right)  }{2L^{2}}\right)  ,
\]
for $\sin(hL/2) \ll L/h$. This results shows that \emph{the
dependence } $J_{c}(\Phi)$ \emph{for junction stacks always has
local maxima at} $\Phi=k\Phi_{0}$, in contrast to the Fraunhofer
dependence for which the critical current vanishes at these points.
To find the critical current behavior in the whole field range in
the region $h/L\gg1$, we numerically found maximum of $J(\beta)$
with respect to $\beta$ and different $hL=2\pi \Phi/\Phi_{0}$ and
$L$. Figure \ref{Fig-Jc-h-small} illustrates the field dependence of
critical current and current dependence of the lattice structure
within one oscillation period $2.5\Phi_{0}<\Phi<3.5\Phi_{0}$ for
$L=2\lambda_{J}$. To visualize the lattice structures, we represent
the values of $|\cos\varphi_{a}|$ by grey level. In most part of the
current-field diagram the rectangular lattice is realized shown by
light grey ($|\cos \varphi_{a}|=1$). The triangular lattice shown by
black ($|\cos\varphi_{a} |=0$) appears in the ground state only in
vicinity of the point $\Phi =3\Phi_{0}$. Exactly at this point the
lattice remains triangular up to the critical current. Slightly away
from this point the lattice deforms with increasing current. In the
range of parameters given by Eq.\ (\ref{Two-minima}) the dependence
$|J(\beta)|$ has two maxima within $0<\beta<\pi/2$ (see left plot in
Fig.\ \ref{Fig-En-J-beta}). As a consequence, the field dependence
of the critical current has kinks related to switching between these
maxima.

\section{Slow dynamics in overdamped regime: Oscillations of the flux-flow
voltage \label{Sec:VoltOsc}}

\begin{figure}[ptb]
\begin{center}
\includegraphics[width=3.2in]{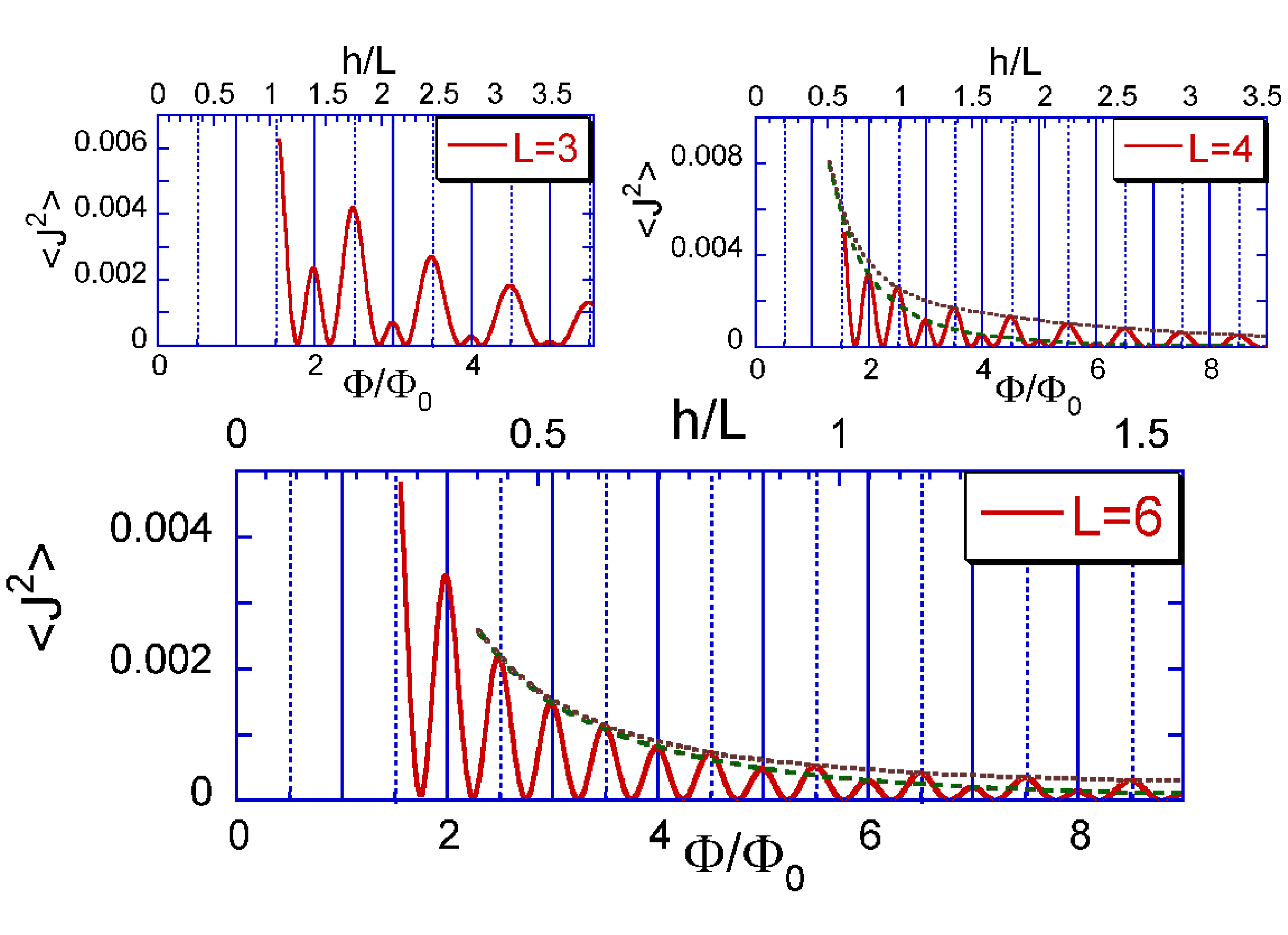}
\end{center}
\caption{The representative field dependences of mean-squared average of
current, $\langle J^{2}(\alpha)\rangle$, with respect to the lattice
displacements which determines the amplitude of relative voltage oscillations
$\delta U/U_{ff}$ at slow velocities via Eq.\ (\ref{deltaU}).}
\label{Fig-J2-h}
\end{figure}\begin{figure}[ptb]
\begin{center}
\includegraphics[width=3.in]{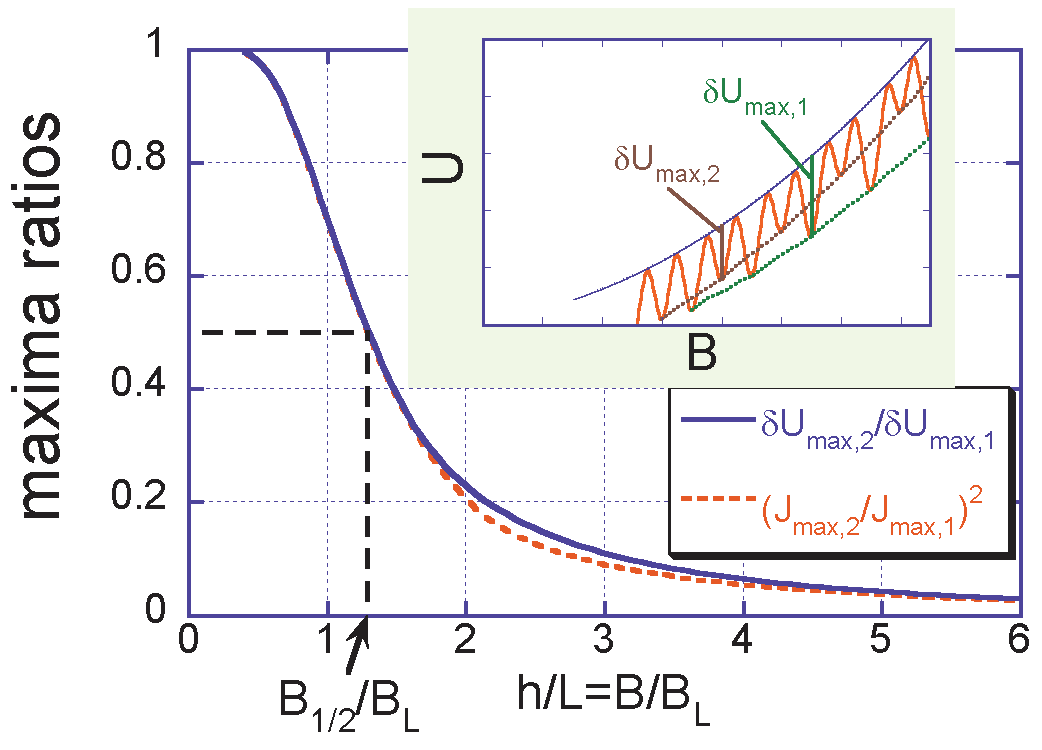}
\end{center}
\caption{The field dependence of the ratio of voltage-oscillation
maxima at integer-flux-quanta points ($\delta U_{max,2}$) and at
half-integer-flux-quanta points ($\delta U_{max,1}$). The inset
illustrates definitions of $\delta U_{max,1}$ and $\delta U_{max,2}$
in the schematic voltage-field dependence. For comparison we also
show plot of the ratio of the critical-current maxima squared
$(J_{max,2}/J_{max,1})^{2}$. Extraction of the typical field
$B_{1/2}$ from the analysis of the voltage oscillations allows
accurately evaluate the anisotropy factor $\gamma$ using Eq.\
(\ref{gammaFromosc}).} \label{Fig-MaxRatio}
\end{figure}

When the external current
flowing across the layers
exceeds the critical current, the lattice starts to move. In
general, dynamic behavior is quite complicated. A simple situation
is realized only at slow lattice motion in the overdamped case when
the lattice deformations have time to adjust to the current lattice
position. In this case the lattice moves in the periodic potential
given by its static energy (\ref{EnSmooth}) and one can use static
results to predict the I-V dependences. On the other hand, one can
expect that the voltage oscillations are less sensitive to
inhomogeneities than the critical-current oscillations, because the
homogeneously moving lattice smears away disorder. The critical
current is also smeared by thermal fluctuations. These are possible
reasons why it is easier to observe and interpret the magnetic
oscillations in the flux-flow resistivity than in the critical
current.\cite{Ooi01,Kakeya05,ZhuPRB05,Urayama06}

In general, the dynamic behavior also depends on the dissipation
mechanism. In BSCCO in a wide range of magnetic fields the flux-flow
resistivity is mainly determined by the in-plane quasiparticle
conductivity $\sigma_{ab} $.\cite{JVflux-flow} Only when the
magnetic field exceeds a typical value
$B_{\sigma}=\sqrt{\sigma_{ab}/\sigma_{c}}\Phi_{0}/(\sqrt{2}\pi\gamma^{2}
s^{2})$, the c-axis conductivity, $\sigma_{c}$, gives dominating
contribution to the flux-flow dissipation. In this limit the
flux-flow resistivity becomes field-independent. In BSCCO the field
$B_{\sigma}$ is typically several times larger than the crossover
field $B_{\mathrm{cr }}$. An important feature of the in-plane
dissipation regime at $B<B_{\sigma}$ is that the lattice velocity at
fixed applied current is very sensitive to lattice structure, the
smallest velocity is realized for the triangular lattice and the
largest velocity is realized for the rectangular lattice. As the
lattice structure in the regime $B\gtrsim B_{L}$ continuously
changes with lattice displacement, the dynamic behavior in this
regime is rather complicated. To avoid this complications, we limit
ourself here by a simple case of dominating c-axis dissipation in
the crossover region, $B_{L}>B_{\sigma}$. In this case the
viscous-friction coefficient weakly depends on lattice structure.

In the case of structure-independent viscous-friction coefficient $\nu_{ff}$,
time variation of the lattice phase shift obeys equation
\begin{equation}
\nu_{ff}\frac{d\alpha}{dt}+J(\alpha)=J_{\mathrm{ext}}, \label{TimeEq}
\end{equation}
where $J_{\mathrm{ext}}$ is the external current, the current $J(\alpha)\equiv
J(\alpha,hL,h)$ is given by Eq.\ (\ref{JosCurrSmooth}) (for brevity we again
skip in equations its dependence on the magnetic field and size), and the
viscosity coefficient, $\nu_{ff}$, is related to the flux-flow resistance of
the stack, $R_{ff}$,
\[
\nu_{ff}=\frac{N\Phi_{0}}{2\pi cR_{ff}}.
\]
where $N$ is the number of junctions in the stack. The voltage drop per one
junction $U$ is related to $d\alpha/dt$ by the Josephson relation
\[
U=\frac{\Phi_{0}}{2\pi c}\frac{d\alpha}{dt}.
\]
Solution of Eq.\ (\ref{TimeEq}) is given by the implicit relation
\[
\int_{0}^{\alpha}\frac{\nu_{ff}d\alpha^{\prime}}{J_{\mathrm{ext}}
-J(\alpha^{\prime})}=t,
\]
from which we obtain the average phase change rate,
\[
\overline{\frac{d\alpha}{dt}}=\left[  \frac{1}{\pi}\int_{0}^{\pi}\frac
{\nu_{ff}d\alpha}{J_{\mathrm{ext}}-J(\alpha)}\right]  ^{-1},
\]
and the flux-flow voltage
\begin{equation}
\frac{U}{U_{ff}}=\left[  \frac{J_{\mathrm{ext}}}{\pi}\int_{0}^{\pi}
\frac{d\alpha}{J_{\mathrm{ext}}-J(\alpha)}\right]  ^{-1} \label{FluxFlowU}
\end{equation}
with $U_{ff}=R_{ff}J$ being the bare flux-flow voltage without the
periodic potential. As the current $J(\alpha)\equiv J(\alpha,hL,h)$
oscillates with the magnetic field, this flux-flow voltage will also
experience similar field oscillations. In particular, when the
external current significantly exceeds the critical current,
$J_{\mathrm{ext}}\gg J_{c}$, we obtain weak oscillating correction
to the flux-flow voltage, $\delta U=U-U_{ff}$,
\begin{equation}
\delta U/U_{ff}\approx-\langle J^{2}(\alpha)\rangle/J_{\mathrm{ext}}^{2},
\label{deltaU}
\end{equation}
where $\langle f(\alpha)\rangle\equiv(1/\pi)\int_{0}^{\pi} f(\alpha)d\alpha$
is the average with respect to the lattice phase shift. As $\delta
U\propto\left\langle J^{2}(\alpha)\right\rangle $, the behavior of $\delta U$
is overall similar to the behavior of the critical current but the amplitude
of voltage oscillations roughly scales as the critical current squared. Figure
\ref{Fig-J2-h} shows the field dependences of the average $\langle
J^{2}(\alpha)\rangle$, which determines the amplitude of weak voltage
oscillations, for three junctions sizes, $L=3$, 4, and 6.

Consider in more details behavior of $\delta U$ at the points $\Phi
=(j/2)\Phi_{0}$. To find the amplitude of the small voltage
correction, $\delta U_{max,1}$, at the half-integer flux quanta
points, $\Phi =(k+1/2)\Phi_{0}$, we have to find the
$\alpha$-average of $J_{1}^{2}(\alpha)$ where the current
$J_{1}(\alpha)$ is given by Eq.\ (\ref{Curr-half-int}) with the
parameter $\tilde{m}$ given by Eq.\ (\ref{half-int-m}). Similarly,
the amplitude of the voltage oscillation at the points
$\Phi=k\Phi_{0}$, which we notate as $\delta U_{max,2}$, is
determined by $\langle J_{2}^2(\alpha )\rangle$ where the current
$J_{2}(\alpha)$ is given by Eq.\ (\ref{Curr-int}) with the parameter
$m$ given by Eq.\ (\ref{int-m}). Figure \ref{Fig-MaxRatio} shows the
computed field dependence of the ratio $\delta U_{max,2}/\delta
U_{max,1}$ together with the ratio $(J_{max,2} /J_{max,1})^{2}$. In
practice, to extract the ratio $\delta U_{max,2}/\delta U_{max,1}$
from experimental voltage-field dependence, one should plot smooth
curves via local maxima and two sets of local minima as it is
illustrated in the inset of Fig.\ \ref{Fig-MaxRatio}, subtract the
two minima curves from the maxima curve, and compute the ratio of
the differences. One can see that the field dependences of $\delta
U_{max,2}/\delta U_{max,1}$ and $(J_{max,2} /J_{max,1})^{2}$ are
almost identical. This plot gives possibility for accurate
determination of the anisotropy factor from the voltage oscillations
using the field scale. In particular, the ratio $\delta U_{2}/\delta
U_{1}$ drops to $0.5$ at the field $B_{1/2}=1.302B_{L}=1.302
\Phi_{0} L/(2\pi \gamma^{2}s^{3}$) (see Fig.\ \ref{Fig-MaxRatio})
meaning that $\gamma$ can be extracted from this field as
\begin{equation}
\gamma\approx330\sqrt{\frac{L[\mu m]}{B_{1/2}[T]}}. \label{gammaFromosc}
\end{equation}
where we used the BSCCO interlayer spacing $s\approx 1.56$ nm. For
example, using data reported in Ref.\ \onlinecite{Kakeya05} for the
sample H55 with $L=5.5$ $\mu$m, we estimate $B_{1/2}\approx3.6$T
giving a very reasonable estimate $\gamma\approx 408$. This estimate
is significantly larger than the value $\gamma\approx110 $ obtained
by the authors themselves and the difference comes from the value of
the numerical constant in the crossover field.

We have to mention that an alternative mechanism of
$\Phi_{0}/2$-periodic voltage oscillations at fixed current exists
at high lattice velocities due to switching between the Fiske
steps.\cite{UstinovPedersenPRB07} However, experimentally, the most
regular voltage oscillations are observed at voltages much smaller
than the first Fiske
voltage.\cite{Ooi01,Kakeya05,ZhuPRB05,Urayama06}

\section{Summary}

In summary, we studied magnetic oscillations of the critical current
and lattice configurations in stacks of intrinsic Josephson
junctions, which are realized in mesas fabricated from layered
high-temperature superconductors. Depending on the stack lateral
size, oscillations may have either the period of half flux quantum
per junction (wide-stack regime) or one flux quantum per junction
(narrow-stack regime). We studied in detail the crossover between
these two regimes. Typical size separating the regimes is
proportional to the magnetic field meaning that the crossover can be
driven by the magnetic field. In the narrow-stack regime the lattice
structure experiences periodic series of first-order phase
transitions between aligned rectangular configuration and triangular
configuration. The triangular configurations in this regime is
realized only in narrow regions near magnetic-field values
corresponding to integer number of flux quanta per junction. For
slow lattice motion similar crossover can also be observed in the
oscillations of the flux-flow resistivity. Quantitative study of the
crossover allows for a very accurate evaluation of the anisotropy
factor.

\section{Acknowledgements}

The author thanks I.\ Kakeya, Y. Latyshev, and T.\ Hatano for useful
discussions of experimental data and to L. Bulaevskii for numerous
discussion of related theoretical issues. This work was supported by
the U.\ S.\ DOE, Office of Science, under contract \#
DE-AC02-06CH11357.

\appendix
\section{Regions of monotonic same-sign solutions for $v(u)$
\label{App:same-sign}}

One can distinguish two types of monotonic solutions depending on
wether or not the smooth phase $v(u)\!=\!(\varphi(u)\!-\!\pi/2)/2$
changes sign inside the junction (see Fig.
\ref{Fig-SolutionDiagram}). For the changing-sign solution the
condition $0\!<\!m\!<\!1$ always holds. In this Appendix we find the
boundary values of $\alpha$ separating these two types of monotonic
solutions. For definiteness, we consider the region
$2k\pi\!<\!hL\!<\!(2k+1)\pi$ and
$-\pi/2+hL-2k\pi\!<\!\alpha\!<\!\pi/2$ (grey area in the lower part
of the phase diagram in Fig.\ \ref{Fig-SolutionDiagram}). For the
monotonic changing-sign solution in this range we have
$0\!<\!\varphi_{0}\!<\!\pi/2$, $\pi/2\!<\!\phi_{L}\!<\!\pi$, i.e.,
\begin{align*}
\varphi_{0}  &  =\arcsin\sqrt{\frac{1/m}{1+2\cos^{2}\left(
\alpha\right)  }
},\\
\varphi_{L}  &  =\pi-\arcsin\sqrt{\frac{1/m}{1+2\cos^{2}\left(
hL-\alpha \right)  }}.
\end{align*}
There are two boundaries in the region, one corresponding to the
condition $\phi_{0}=\pi/2$ ($v_{0}=0$) below $\alpha=\pi/2$ and
another corresponding to $\phi_{L}=\pi/2$ ($v_{L}=0$) above
$\alpha=-\pi/2+hL-2k\pi$ (see Fig. \ref{Fig-SolutionDiagram}). For
the first boundary, $\alpha_{0}(h,L)$, from Eqs.\
(\ref{sin_phi0_mon}) and (\ref{Eq_m_monot}) we obtain
\begin{subequations}
\begin{align}
\sqrt{m_{0}}  &  \left[  K(m_{0})\!-\!F\left(  \arcsin\sqrt{\frac{1/m_{0}
}{1\!+\!2\cos^{2}(hL\!-\!\alpha_{0}) }},m_{0}\right)  \right] \nonumber\\
&  =\frac{\sqrt{8}L}{h},\\
m_{0}  &  =\frac{1}{1+2\cos^{2}\left(  \alpha_{0}\right)  },
\end{align}
\end{subequations}
where $K\left(  m\right) = F(\pi/2,m)$ is the complete elliptic
integral of the first kind \cite{Abramovitz} and we used the
identity $F\left(\pi -\beta,m\right)\!-\!K(m)\!=\!K(m)\!-\!F\left(
\beta,m\right)  $. Analyzing similar equation for the second
boundary, $\alpha_{L}(h,L)$, we find that it is related to
$\alpha_{0}(h,L)$ as $\alpha_{L}(h,L)=hL-2k\pi-\alpha_{0}(h,L)$. One
can check that $\alpha_{0}(h,L)\rightarrow\pi/2$ for $hL\rightarrow
2k\pi,(2k+1)\pi$ and for $L\rightarrow\infty$, i.e., the region of
the same-sign monotonic solution vanish in these limits.

\section{Weak finite-size effects at large $L$ \label{App:LargeL}}

In this Appendix we derive finite-size corrections to the critical
current due to interaction between the surface solitons. Consider
for definiteness the case of monotonic solution. The nonmonotonic
solution can be treated similarly. In the limit $L/h\gg1$ the
parameter $m$ in Eq.\ (\ref{MonotRelat}) is close to one. Separating
small correction, $m=1-\eta/2$ with $\eta\ll1$, we evaluate the
integral as
\[
\int_{v_{0}}^{v_{L}}\frac{dv}{\sqrt{1+\eta/2-\cos^{2}\left(
2v\right)  } }\approx\frac{1}{2}\ln\left(  -\frac{32}{\eta}\tan
v_{0}\tan v_{L}\right)  .
\]
This gives the following result for $\eta$
\begin{equation}
\eta\approx-32\tan v_{0}\tan v_{L}\exp\left(  -\sqrt{8}L/h\right)
,\label{FinSize-EllParCorr}
\end{equation}
corresponding to the elliptic-function parameter $m\approx1+16\tan
v_{0}\tan v_{L}\exp\left(  -\sqrt{8}L/h\right)  $. The boundary
conditions can be represented as
\begin{align*}
\cos\left(  2v_{0}\right)   &
\approx\frac{1+\eta/4}{\sqrt{1+2\cos^{2}\left(
\alpha\right)  }},\\
\cos\left(  2v_{L}\right)   &
\approx\frac{1+\eta/4}{\sqrt{1+2\cos^{2}\left( hL-\alpha\right)  }}.
\end{align*}
One can neglect shift of $\eta$ due to the finite-size corrections
to $v_{0}$ and $v_{L}$. Without interaction between edge
deformations, the energy can be written as\cite{AEKPRB02}
\begin{align*}
\mathcal{E}_{0} &  =\frac{1}{h}\sin\left(  2v_{00}\right)
\cos\left( \alpha\right)  -\frac{1}{h}\sin\left(  2v_{L0}\right)
\cos\left(
hL-\alpha\right)  \\
&  +\frac{1}{h}\int_{v_{00}}^{v_{L0}}dv\sqrt{1-\cos\left(  4v\right)  }
-L\frac{2}{2h^{2}}\\
&  =\frac{1}{h\sqrt{2}}\left(
2\!-\!\sqrt{2\!+\!\cos2\alpha}\!-\!\sqrt {2\!+\!\cos2\left(
\alpha\!-\!hL\right)  }\right)  \!-\!\frac{L}{h^{2}}.
\end{align*}
where $v_{00}$ and $v_{L0}$ are the surface deformations neglecting
finite-size correction,
\begin{align*}
\tan v_{00} &  =\frac{1-\sqrt{1+2\cos^{2}\alpha}}{\sqrt{2}\cos\alpha},\\
\tan v_{L0} &  =-\frac{1-\sqrt{1+2\cos^{2}\left(  \alpha-hL\right)
}} {\sqrt{2}\cos\left(  \alpha-hL\right)  }.
\end{align*}
The finite-size correction to the energy change can now be estimated
as
\begin{align*}
\delta\mathcal{E} &
\!=\!\frac{1}{h}\int_{v_{00}}^{v_{L0}}\!dv\left(
\sqrt{1\!+\!\eta\!-\!\cos\left(  4v\right)
}\!-\!\sqrt{1\!-\!\cos\left(
4v\right)  }\right)  \!-\!\frac{L\eta}{2h^{2}}\\
&  \approx\frac{\eta}{h\sqrt{32}}.
\end{align*}
Therefore, the total total energy can be written as
\begin{align}
&  \mathcal{E}(\alpha,h,hL)=-\frac{L}{h^{2}}\nonumber\\
&  +\frac{1}{\sqrt{2}h}\left(
2-\sqrt{2+\cos(2\alpha)}-\sqrt{2+\cos\left[
2\left(  \alpha-hL\right)  \right]  }\right)  \nonumber\\
&  +\frac{8\sqrt{2}}{h}\frac{\cos(\alpha)\cos\left(
\alpha-hL\right) \exp\left(  -\sqrt{8}L/h\right)  }{\left(
1\!+\!\sqrt{2\!+\!\cos(2\alpha )}\right)  \left(
1\!+\!\sqrt{2\!+\!\cos\left[  2(\alpha\!-\!hL)\right]
}\right)  .}\label{FinSizeEnTotal}
\end{align}
The last correction term describes the exponentially small
interaction energy between the surface solitons. It is important to
note that the finite-size correction breaks $\pi$-periodicity with
respect to parameter $hL$ meaning that the states with
$\Phi=k\Phi_{0}$ and $\Phi=(k+1/2)\Phi_{0}$ are not equivalent any
more.

For the Josephson current we obtain \begin{widetext}
\begin{align*}
J(\alpha,h,hL)  &  =\frac{1}{\sqrt{2}h}\left(
\frac{\sin2\alpha}{\sqrt {2+\cos2\alpha}}+\frac{\sin2\left(
\alpha-hL\right)  }{\sqrt{2+\cos2\left(
\alpha-hL\right)  }}\right) \\
&  +\frac{8\sqrt{2}}{h}\exp\left(  -\frac{\sqrt{8}L}{h}\right) \frac
{\partial}{\partial\alpha}\left[  \frac{\cos\alpha\cos\left( \alpha
-hL\right)  }{\left(  1+\sqrt{2+\cos2\alpha}\right)  \left( 1+\sqrt
{2+\cos2\left(  \alpha-hL\right)  }\right)  }\right].
\end{align*}
Assuming that without the finite-size correction the maximum current
flows at $\alpha=\alpha_{m}(hL)$, we obtain for the finite-size
correction to the critical current
\[
\delta J_{c}(h,hL)=\frac{8\sqrt{2}}{h}\exp\left(
-\frac{\sqrt{8}L}{h}\right) \frac{\partial}{\partial\alpha}\left[
\frac{\cos\alpha\cos\left( \alpha-hL\right)  }{\left(
1+\sqrt{2+\cos(2\alpha)}\right)  \left( 1+\sqrt{2+\cos\left[ 2\left(
\alpha-hL\right)  \right]  }\right)  }\right]
_{\alpha=\alpha_{m}(hL)}.
\]
\end{widetext}
In particular, near the maxima $hL=\pi j$ ($\Phi=j\Phi_{0}/2$),
using the result \cite{AEKPRB02} $\alpha_{m}(0)=0.921$, we obtain
result (\ref{CritCurrMaxLargeL}).


\begin{thebibliography}{99}
\bibitem{note-length}Note that in the previous paper \cite{AEKPRB02} the
typical length scale $l_{B}$ has been defined with different
numerical factor, $l_{B}=L_{B}/2\sqrt{2}$. This earlier definition
was based on the decay length of the surface deformation. A more
detailed study of this paper shows that the crossover between the
stack regimes is better described by the condition $L\sim L_{B}$.
\end{thebibliography}
\end{document}